\documentclass[a4paper,12pt]{article}
\pdfoutput=1
\usepackage[usenames]{color}
\usepackage[colorlinks=true
,urlcolor=blue
,anchorcolor=blue
,citecolor=green
,filecolor=blue
,linkcolor=blue
,menucolor=blue
,pagecolor=blue
,linktocpage=true
]{hyperref}
\usepackage[centertags]{amsmath}
\usepackage{amsfonts}
\usepackage{amssymb}
\usepackage{bbm}
\usepackage{amsthm}
\usepackage{newlfont}
\usepackage[a4paper]{geometry}
\usepackage{graphicx}
\usepackage{dsfont}
\usepackage{feynmf}
\usepackage{cleveref}
\usepackage{array}

\DeclareMathOperator{\trace}{Tr}

\newcommand{\tlambda}{\tilde{\lambda}}
\newcommand{\hlambda}{\hat{\lambda}}
\newcommand{\atan}{\arctan}
\newcommand{\YM}{\mathrm{YM}}
\newcommand{\fer}{\mathrm{fer.}}

\newcommand{\dho}{\partial}

\newcommand{\olra}{\overleftrightarrow}

\newcommand{\ed}{\,.}
\newcommand{\ec}{\,,}
\newcommand{\ecq}{\ec\quad}

\newcommand{\cA}{\ensuremath{\mathcal{A}}}

\newcommand{\cD}{\ensuremath{\mathcal{D}}}

\begin{document}

\title{
\vspace{-20mm}
\hfill{\small \tt WIS/13/12-JUL-DPPA}\vskip 5pt
Correlation Functions of Large $N$ Chern-Simons-Matter Theories and Bosonization in Three Dimensions}
\author{Ofer Aharony, Guy Gur-Ari, and Ran Yacoby\\\\
{\it Department of Particle Physics and Astrophysics}\\
{\it Weizmann Institute of Science, Rehovot 76100, Israel}\\
{\small{\tt e-mails~:~Ofer.Aharony,~Guy.GurAri,~Ran.Yacoby@weizmann.ac.il}}}

\maketitle
\vspace{-7mm}
\begin{abstract}
We consider the conformal field theory of $N$ complex massless scalars in $2+1$ dimensions, coupled to a $U(N)$ Chern-Simons theory at level $k$. This theory has a 't Hooft large $N$ limit, keeping fixed $\lambda \equiv N/k$. We compute some correlation functions in this theory exactly as a function of $\lambda$, in the large $N$ (planar) limit. We show that the results match with the general predictions of Maldacena and Zhiboedov for the correlators of theories that have high-spin symmetries in the large $N$ limit. It has been suggested in the past that this theory is dual (in the large $N$ limit) to the Legendre transform of the theory of fermions coupled to a Chern-Simons gauge field, and our results allow us to find the precise mapping between the two theories. We find that in the large $N$ limit the theory of $N$ scalars coupled to a $U(N)_k$ Chern-Simons theory is equivalent to the Legendre transform of the theory of $k$ fermions coupled to a $U(k)_N$ Chern-Simons theory, thus providing a bosonization of the latter theory. We conjecture that perhaps this duality is valid also for finite values of $N$ and $k$, where on the fermionic side we should now have (for $N_f$ flavors) a $U(k)_{N-N_f/2}$ theory. Similar results hold for real scalars (fermions) coupled to the $O(N)_k$ Chern-Simons theory.
\end{abstract}

\newpage

\tableofcontents

\setlength{\unitlength}{1mm}

\section{Introduction and Summary of Results}

Two of the simplest conformal field theories in $2+1$ dimensions are the theory of $N$ real massless free scalars $\varphi^i$, in the sector of operators that are singlets of $O(N)$, and the ``critical $O(N)$ model''. The latter may be viewed as the IR limit of the deformation of the free theory by $g_4 (\varphi^i \varphi^i)^2$ (when fine-tuning the IR scalar mass to zero), or equivalently by adding an auxiliary field $\sigma$ and deforming the free theory by $\sigma \varphi^i \varphi^i$ (and fine-tuning a linear term in $\sigma$ to get to a non-trivial fixed point). Due to the latter description we will call the ``critical $O(N)$ model'' the Legendre transform of the free theory\footnote{The two theories are not precisely Legendre transforms, except in the large $N$ limit, since we need to flow to the IR and to fine-tune to get to the fixed point. However, we will use this name for this type of relation between theories, for lack of a better name.}.

Two other simple conformal field theories in $2+1$ dimensions are the theory of $N$ massless free fermions $\psi^a$ (again limited to singlets of $O(N)$), and the Gross-Neveu model. One can formally define the latter by deforming the free fermion theory by $(\psi^a \psi^a)^2$, but since this operator is not renormalizable this definition does not make much sense (except in the large $N$ limit). As above, one can also formally define it by performing a Legendre transform with respect to the operator $(\psi^a \psi^a)$. One way to properly define the Gross-Neveu theory is as a theory that has a scalar operator $\hat\sigma$ (analogous to the auxiliary field $\sigma$ above), such that when we deform the theory by $({\hat\sigma}^2)$ it flows to the free fermion theory (when fine-tuning the IR fermion mass to zero).

The free theories described above both have high-spin symmetries of every even spin $s=2,4,6,\cdots$, generated by conserved currents $J^{(s)}$. In the interacting theories these symmetries (for $s > 2$) are broken, though the effect of the breaking is small in the large $N$ limit\footnote{The divergences of the high-spin currents and their anomalous dimensions are proportional to $1/N$, but some correlators feel the breaking even in the large $N$ limit.}. In this limit all local operators in these theories are products of the high-spin currents $J^{(s)}$ and of one additional scalar operator $J^{(0)}$ (given by $(\varphi^i \varphi^i)$ in the bosonic case, and by $(\psi^a \psi^a)$ in the fermionic case).

The two bosonic theories are simply related by a Legendre transform with respect to $J^{(0)}$, as are the two fermionic theories, but at first sight there is no relation between the bosonic and fermionic theories. As far as we know, the first hint for such a relation came by considering their gravity duals. All the theories discussed above have a good $1/N$ expansion (see \cite{Moshe:2003xn} for a review), so it is natural to suggest that they could have classical gravitational duals (by the AdS/CFT correspondence \cite{Maldacena:1997re,Gubser:1998bc,Witten:1998qj}) at large $N$, living on $AdS_4$. These duals should have massless high spin fields, to match with the field theory spectrum. Indeed, it was suggested in \cite{Klebanov:2002ja,Sezgin:2003pt} (see also \cite{Sundborg:2000wp,Witten_talk,Sezgin:2002rt}) that the bosonic theories are dual to the type A high-spin gravity theories of Vasiliev \cite{Fradkin:1987ks}, and the fermionic theories to the type B high-spin gravity theories of Vasiliev. In the gravity language, the difference between the two bosonic (fermionic) theories is just a different choice of boundary conditions for the bulk scalar field which is dual to $J^{(0)}$; in the classical gravity limit this is equivalent to a Legendre transform \cite{Witten:2001ua,Gubser:2002vv,Petkou:2003zz}. Strong evidence for this equivalence was found in \cite{Giombi:2009wh,Giombi:2010vg}, and suggested derivations of the equivalence were presented in \cite{Das:2003vw,Douglas:2010rc,Koch:2010cy,Jevicki:2011ss}.

The type A and type B high-spin gravity theories mentioned above have the same spectrum but different interactions. However, there is a family of high-spin gravity theories labeled by a parameter $\theta$ (appearing in the Vasiliev equations of motion) that interpolates between the type A and type B theories, such that the bosonic theories arise when $\theta=0$ and the fermionic ones when $\theta=\pi/2$. This hints that perhaps there is a family of field theories (that is continuous, at least in the large $N$ limit) which interpolates between the bosonic and fermionic theories described above. The theories with $\theta\neq 0,\pi/2$ are not parity-invariant. It was conjectured in \cite{Giombi:2011kc} that they arise by coupling the bosonic/fermionic theories to $O(N)$ Chern-Simons theories at level $k$. For infinite $k$ this just implements the reduction to singlets of $O(N)$, but for finite $\lambda \equiv N/k$ in the large $N$ limit it provides a parity-breaking modification of the bosonic and fermionic theories, by a parameter that is continuous in the large $N$ limit (though it is discrete for finite $N$). This conjecture was tested in \cite{Giombi:2011kc,Aharony:2011jz}, where it was shown that the spectrum of operators in these theories is independent of $\lambda$ in the large $N$ limit. This is consistent with the fact that the high-spin fields in the dual gravity theory are classically massless; presumably they acquire a mass (whenever the field theory is not free) at order $1/N$ by loop corrections in the bulk \cite{Girardello:2002pp}.

Significant support for this conjectured relation between the bosonic and fermionic theories was recently given in \cite{Maldacena:2011jn,Maldacena:2012sf}, where it was shown that the large $N$ correlation functions in these theories could be computed just by knowing that they have a high-spin symmetry that is broken by $1/N$ effects (plus some additional technical assumptions). Maldacena and Zhiboedov showed that these correlation functions could be expressed in terms of two (or three) parameters appearing in the non-conservation equations of the high-spin currents. It is natural to assume that these parameters map to the parameters $N$ and $k$ mentioned above, but the precise relation is not known (when there is a third parameter it can be identified with the coefficient of an extra $g_6 (\varphi^i \varphi^i)^3$ interaction, that is exactly marginal in the large $N$ limit \cite{Aharony:2011jz}). Maldacena and Zhiboedov found that if one starts from the free bosonic theory and turns on the coupling (in their language, turning on the breaking of the high-spin symmetry), then in the limit of infinite coupling one ends up with the correlation functions of the Gross-Neveu model (the Legendre transform of the free fermionic theory). The Legendre-transformed statement is also true: by starting from the correlation functions of the free fermionic theory and turning on the coupling, they found that in the limit of infinite coupling one ends up with the ``critical $O(N)$ model'' (the Legendre transform of the free bosonic theory).

This suggests that for each theory mentioned above, there are actually four different ways to describe it and to compute its large $N$ correlation functions : (a) It can be described as a theory of $N_s$ massless
scalars coupled to a $O(N_s)_{k_s}$ Chern-Simons theory; (b) It can be described as a theory of $N_\fer$ massless fermions coupled to a $O(N_\fer)_{k_\fer}$ Chern-Simons theory; (c) It can be described purely algebraically as a theory with a slightly-broken high-spin symmetry, and its large $N$ correlators can be expressed in terms of the parameters ${\tilde N}$ and $\tilde \lambda$ of \cite{Maldacena:2012sf}; (d) It can be described as Vasiliev's high-spin gravity theory with a parameter $\theta$. The first two descriptions exist also for finite $N$, while in the latter two it is only known how to compute in the planar limit. The discussion above implies that in the large $N$ limit all of these descriptions are equivalent (up to possible Legendre transforms, in particular between the first and second descriptions), but the precise mapping between them is not yet known.

Our main goal in this paper is to clarify the relation between the first three descriptions of these theories, and to compute the precise mapping between their parameters (in the large $N$ limit)\footnote{We will not discuss here the mapping to the gravitational side. The results of \cite{Giombi:2011kc} for even contributions to correlators imply that in this mapping ${\tilde \lambda} = \tan(\theta)$, and this is confirmed by more detailed computations in \cite{Yin}.}. All the $O(N)$ theories discussed above have also $U(N)$ versions, where we start with $N$ complex scalars (fermions) and couple them to a $U(N)_k$ Chern-Simons theory. The $U(N)$ theories have conserved currents $J^{(s)}$ (in the large $N$ limit) also for odd spins $s$, but in the large $N$ limit the correlators of the even spin operators in these theories are equivalent to those of the $O(N)$ theories. In this paper we will perform computations for the $U(N)$ case, since in this case there is a spin-one current that takes a simpler form than the higher spin currents; however our conclusions should be equally valid for the $O(N)$ case.

Most of our paper is devoted to performing exact computations of large $N$ correlation functions in the bosonic theories described above.
After introducing our theories and methods in section \ref{CSconv}, we compute in section \ref{scalars} some exact planar correlation functions of the scalar fields in these theories. These are then used in section \ref{correlators} to compute exactly (as a function of $\lambda$) some planar correlators of gauge-invariant operators in the bosonic theories. Specifically we compute some $2$-point and $3$-point functions of the operators $J^{(s)}$ with $s=0,1,2$.

Section \ref{micro} contains our main results. We begin by matching our results for the theories of scalars coupled to a $U(N)_k$ Chern-Simons theory to the general results of \cite{Maldacena:2012sf}. There are two different standard definitions of the Chern-Simons level. We will denote by $k$ the definition we will use in our computations (which is sometimes called the ``renormalized coupling''; it arises, for instance, from regularization by dimensional reduction). However, defining the level by a regularization using a Yang-Mills term at high energies gives a different (at one-loop) definition of the level, $k_{\YM} = k - N$. Some of our results are simpler to express in the first language and some in the second language; of course the translation between them is straightforward, and we will try to carefully distinguish the two everywhere. The matching to \cite{Maldacena:2012sf} works better using $k$ and $\lambda \equiv N / k$; in these variables we find that the parameters of \cite{Maldacena:2012sf} are related to the bosonic rank and level by\footnote{The same value of ${\tilde \lambda}$ is found in supersymmetric generalizations of these theories, by completely different methods, in \cite{Yin}.}
\begin{equation}
{\tilde N} = 2 N \, \frac{\sin(\pi \lambda)}{\pi \lambda} \, , \qquad {\tilde \lambda} = \tan \left( \frac{\pi \lambda}{2} \right).
\end{equation}
In particular ${\tilde \lambda}$ diverges as $\lambda \to 1$, which corresponds to an infinite coupling ($N/k_{\YM} \to \infty$) using the Yang-Mills regularization, and in this limit we should approach the fermionic theory.

The mapping to the fermionic theory is nicer to describe using the Yang-Mills regularization, which we will use in the rest of this introduction\footnote{We will always use this convention whenever we have $k_{\YM}$ appearing, either in the level or in the rank of our theories.}.
Taking the strong coupling limit, we find that the theory of $N$ scalars coupled to a $U(N)_{k_{\YM}}$ Chern-Simons theory matches (in the large $N$,$k_{\YM}$ limit) to the (Legendre transform of the) theory of $k_{\YM}$ fermions coupled to a $U(k_{\YM})_N$ Chern-Simons theory\footnote{In our computations we do not determine the sign of the Chern-Simons coupling, so the level may also be $(-N)$.}. The two Chern-Simons theories in question are related by level-rank duality \cite{Naculich:1990pa,Camperi:1990dk,Mlawer:1990uv}, and our claim is that coupling one of them to a massless scalar in the fundamental representation is exactly equivalent (up to a Legendre transform) to coupling the other to a massless fermion in the fundamental representation. This may be viewed as a bosonization of the fermionic theory, expressing it purely in bosonic variables. The generalization to the $O(N)$ case is straightforward, mapping the theory of $N$ real scalars coupled to an $O(N)_{k_{\YM}}$ Chern-Simons theory to the (Legendre transform of the) theory of $k_{\YM}$ real fermions coupled to an $O(k_{\YM})_N$ Chern-Simons theory.

Our results show that the bosonic and fermionic theories have the same correlation functions at large $N$, and thus provide strong evidence for their equivalence in this limit. More precisely, we derive this particular matching of parameters just in the strong coupling limit of the bosonic theory, which is the weak coupling limit of the fermionic theory, but it seems natural that it should extend throughout the parameter space; this is confirmed by preliminary computations of exact 2-point and 3-point correlators in the fermionic theory \cite{GY}.
The mapping of the previous paragraph contradicts computations of the thermal free energy presented in \cite{Giombi:2011kc}, which is why it was not found already in \cite{Giombi:2011kc,Maldacena:2012sf}; we discuss in the final section why these computations do not give the correct answers for the free energy at finite $\lambda$.

While so far we only discussed the large $N$ limit, it is natural to
conjecture that perhaps the equivalence between the two theories is valid also at finite $N$ (this is true for the level-rank duality, and also in similar dualities for supersymmetric theories \cite{Giveon:2008zn,Kapustin:2011gh}). If this is not correct, then the scalar and fermion theories would provide two different quantum generalizations of the same classical high-spin gravity theory. Unfortunately, it is hard to test this conjecture, since at least one side of the duality is always strongly coupled, and we do not know how to perform exact computations at finite $N$. All we have so far is a weak test of this finite $N$ duality, by comparing the mass deformations on both sides. This test suggests that at finite $N$ the precise mapping is from a $U(N)_{k_{\YM}}$ bosonic theory to a $U(k_{\YM})_{N-1/2}$ fermionic theory. If we have $N_f$ flavors in the fundamental representation, a natural generalization would map the theory of $N_f$ massless scalars in the fundamental  of $U(N)$ coupled to the $U(N)_{k_{\YM}}$ Chern-Simons theory to the (Legendre transform of the) theory of $N_f$ massless fermions in the fundamental of $U(k_{\YM})$, coupled to the $U(k_{\YM})_{N-N_f/2}$ Chern-Simons theory. It would be interesting to find ways to test this conjecture. In particular it may be interesting to see how the scalars become fermions and vice versa; one can try to analyze this\footnote{We thank N. Itzhaki for this suggestion.} by looking at open Wilson lines ending on scalars/fermions and checking which anyonic statistics the ends of these Wilson lines obey, as a function of $N$ and $k_{\YM}$.

\section{Vector Model with Chern-Simons Interactions}
\label{CSconv}

Consider the theory of a complex scalar field $\phi$ in the fundamental representation of $U(N)$, coupled to gauge bosons $A_\mu$ with Chern-Simons interactions at level $k$ in three Euclidean dimensions. The action is
\begin{align}
  S = S_{\mathrm{CS}} + S_{\mathrm{scalar}} \ed
  \label{eq:L}
\end{align}
The Chern-Simons action $S_{\mathrm{CS}}$ is given by
\begin{align}
  S_{\mathrm{CS}} &= \frac{ik}{4\pi} \int
  \text{Tr}_N\!\left(A\wedge dA + \frac{2}{3}A^3\right)
  = \frac{i k C_1(N)}{4\pi} \int \! d^3x \, \epsilon^{\mu\nu\rho}
  \left(A^a_{\mu}\dho_{\nu}A^a_{\rho}
  + \frac{1}{3}f^{abc}A^a_{\mu}A^b_{\nu}A^c_{\rho}\right) \ed
\label{eq:CSaction}
\end{align}
The trace is taken in the fundamental representation, and we normalize the generators with $C_1(N)=-1/2$\footnote{For a representation $R$ of $U(N)$ with generators $\{T^a\}$, $C_1(R)$ is defined by $\trace_R(T^a T^b)~=~C_1(R) \delta^{ab}$.}.
With this normalization the theory is gauge invariant if $k\in \mathds{Z}$ \cite{Deser:1981wh,Deser:1982vy}. The scalar field action is
\begin{align}
  S_{\mathrm{scalar}} &= \int \! d^3x \left( |\cD_{\mu}\phi|^2 +
  \frac{\lambda_6}{3! N^2} (\phi^\dagger\phi)^3 \right)
  \ec
\end{align}
where $\cD_{\mu} \equiv \dho_{\mu} + A_{\mu}$.

We work in the 't~Hooft large $N$ limit, keeping $\lambda = \frac{N}{k}$ and $\lambda_6$ fixed. In this limit, the theory \eqref{eq:L} is conformal to all orders in perturbation theory in $\lambda$ and $\lambda_6$ \cite{Aharony:2011jz}. In the planar
limit, our theory is closely related to the $O({\hat N})$ theory of $\hat N$ real scalar fields
coupled to an $O({\hat N})$ Chern-Simons theory at level $\hat k$; in this limit the latter theory (with
${\hat N}=2N$) is simply a projection of our theory, keeping only some of its operators. For finite $N$ the two theories are not
equivalent, but all of our large $N$ computations (except for the ones that involve operators that
are projected out when going to $O({\hat N})$) can easily be translated into computations in the
$O({\hat N})$ theory as well.

Let us define light-cone coordinates by $x^{\pm} = x_{\mp} = (x^1\pm ix^2)/\sqrt{2}$. We work in light-cone gauge, $A_- = 0$\footnote{In Euclidean space $A_+ = \bar{A}_-$, but we keep $A_+ \ne 0$. One can think of this prescription as an analytic continuation of light-cone gauge in Minkowski space.}. With this choice of gauge, the $A\wedge A \wedge A$ self-interaction of the gauge field vanishes, and the seagull term $\phi^2 A_{\mu}^2$ is also simplified as we shall see. This greatly reduces the complexity of perturbative calculations. The utility of light-cone gauge in theories of this kind was first noticed in \cite{Giombi:2011kc}.

In this gauge the gluon propagator is
\begin{align}
  \langle  A_\mu^a(-p) A_\nu^b(q) \rangle  &= G_{\nu\mu}(p) \delta^{ab}
  \cdot (2\pi)^3 \delta^3(q-p) \ec \notag \\
  G_{+3}(p) &= - G_{3+}(p)
  = \frac{4\pi i}{k} \frac{1}{p^+} \ec
  \label{gluon-prop}
\end{align}
and the other components of $G_{\nu\mu}$ vanish. At leading order in the 't Hooft large $N$ limit, where scalar loops can be ignored, the gluon propagator in this gauge does not receive corrections, because the $A\wedge A\wedge A$ interaction vanishes and the ghosts are decoupled.

To regulate the theory, we use dimensional regularization in the direction $x^3$, and a cutoff $\Lambda$ on the momentum in the 1-2 plane. The cutoff regulator breaks Lorentz invariance (as does our choice of gauge), conformal invariance and gauge invariance. However, we only encounter power-law divergences, and the counter-terms used to subtract those divergences are completely fixed by demanding that the continuum theory is conformally invariant (in general these counter-terms are not gauge-invariant or Lorentz-invariant). The fact that our final results for the correlation functions of gauge-invariant operators are consistent with the analysis of \cite{Maldacena:2012sf} (see section \ref{micro}) gives strong evidence that Lorentz and gauge invariance are restored in the continuum limit.

The value of $k$ in Chern-Simons theories depends on the regularization; as in any other theory, coupling constants in different regularizations are not the same at higher orders in perturbation theory. In particular, as shown in \cite{Chen:1992ee}, the value of $k$ in our regularization (which is sometimes called the ``renormalized coupling'') differs from the value of $k$ using a regularization involving a Yang-Mills term in the UV by $k_{\mathrm{us}} = k_{\YM} + N = k_{\YM} (1 + \lambda_{\YM})$ (where we define $\lambda_{\YM} \equiv N / k_{\YM}$)\footnote{Without loss of generality we assume $k > 0$. A parity transformation takes $k \to -k$, $\lambda \to -\lambda$.}. In terms of the 't Hooft coupling we have $\lambda_{\mathrm{us}} = \lambda_{\YM} / (1 + \lambda_{\YM})$, so that the two couplings agree perturbatively, but the maximal value of $\lambda$ that can be achieved in the Yang-Mills regularization is $\lambda=1$ (which corresponds to $\lambda_{\YM} \to \infty$, or $k_{\YM} \to 0$).

In the large $N$ limit, the spectrum of operators of the theory \eqref{eq:L} includes a single primary operator $J^{(s)}$ for each integer spin $s \ge 0$, with conformal dimension $\Delta=s+1+O(1/N)$ \cite{Aharony:2011jz}. Each $J^{(s)}$ can be written as a symmetric, traceless tensor that is schematically given by
\begin{align}
  J_{\mu_1\dots\mu_s}^{(s)} = \phi^\dagger_i \cD_{\mu_1} \cdots \cD_{\mu_s}
  \phi^i + \cdots \ed
\end{align}
All other primaries are products of these ``single-trace'' operators. The currents $J_{\mu}\equiv J^{(1)}_{\mu}$, $T_{\mu\nu}\equiv J^{(2)}_{\mu\nu}$ correspond to the unbroken $U(1)$ and Poincar\'e symmetries and are conserved\footnote{Note that the naive $U(1)$ global symmetry acting on the complex scalar fields is gauged, but the equations of motion of the gauge field imply that the symmetry generated by the topologically-conserved current $J^{(1)} = * {\mathrm{tr}}(dA)$, acts on the scalar fields in the same way as the naive $U(1)$ global symmetry, up to an overall factor.}. The currents with $s>2$ are generally not conserved when $\lambda \ne 0$. In this work we will need the explicit form of the following operators :
\newcommand{\leftD}{\overleftarrow{\cD}}
\newcommand{\rightD}{\overrightarrow{\cD}}
\begin{align}
  J^{(0)} &= \phi^\dagger \phi \ec \notag\\
  J_\mu &= i \phi^\dagger \left( \leftD_\mu - \rightD_\mu \right) \phi \ec\notag\\
  T_{\mu\nu} &= \phi^\dagger \left[
  \frac{3}{2} \leftD_{(\mu} \rightD_{\nu)}
  - \frac{1}{4} \rightD_{(\mu} \rightD_{\nu)}
  - \frac{1}{4} \leftD_{(\mu} \leftD_{\nu)}
  \right] \phi
  + \delta_{\mu\nu} \left(\cdots\right) \ed
  \label{currents}
\end{align}
Here $\overrightarrow{\cD}_\mu = \overrightarrow{\partial}_\mu + A_\mu$, $\overleftarrow{\cD}_\mu = \overleftarrow{\partial}_\mu - A_\mu$, and parentheses around indices denote averaging over symmetric permutations. The trace terms in $T_{\mu\nu}$ will not be important for us in this paper. The currents in \eqref{currents} are canonically normalized, namely, the charges $Q\equiv \int d^2x J_0$, $P_\mu \equiv \int d^2x T_{0\mu}$ obey $[Q,\phi^i]=\phi^i$, $[P_\mu,\phi^i]=-i\partial_\mu \phi^i$.

For completeness we now review the results of \cite{Maldacena:2012sf} which will be needed in section \ref{micro}.

\subsection{Review of the Results of Maldacena and Zhiboedov}
\label{mzreview}

Our theory belongs, in the large $N$ limit, to the general class of conformal field theories studied in \cite{Maldacena:2012sf}, which were named ``quasi-boson'' theories. In general, a CFT belongs to this class of theories if it has a large $N$ expansion, and its large $N$ spectrum of operators includes conserved high-spin currents of even spins and a dimension one scalar operator. The large $N$ expansion parameter, denoted by $\tilde{N}$, was defined in \cite{Maldacena:2012sf} to be proportional to the two-point function of the energy-momentum tensor, and the proportionality constant was fixed by requiring that $\tilde{N}=1$ for a free real boson. The theory described in this section, and the Gross-Neveu model coupled to $U(N)_k$ Chern-Simons interactions, are particular examples of ``quasi-boson'' theories.

A closely related class of theories, called ``quasi-fermion'' theories, was defined in \cite{Maldacena:2012sf} by the same properties that were used to define the ``quasi-boson'' theories, except that the scalar primary $J^{(0)}$ has dimension $2+O(1/N)$. Following \cite{Maldacena:2012sf} we will denote this scalar by $\tilde{J}^{(0)}$ to avoid confusion with the quasi-boson case. The Legendre transform of the theory described in this section, and the theory of $N$ free fermions coupled to $U(N)_k$ Chern-Simons interactions, are both examples of ``quasi-fermion'' theories.

The conservation of high-spin currents in either the quasi-boson or quasi-fermion theories is generally violated by double-trace and triple-trace operators. For example, the non-conservation of the $s\!=\!4$ current in the quasi-boson theory takes the schematic form (with appropriate derivatives on the right-hand side)
\begin{align}
\partial \cdot J^{(4)} = a_2 J^{(2)} J^{(0)} + a_3 \left( J^{(0)} J^{(0)} J^{(0)} + J^{(2)} J^{(0)} J^{(0)} \right) \ec
\label{dJ4}
\end{align}
and a similar equation containing only the double-trace term applies in the quasi-fermion case.
Using \eqref{dJ4} Maldacena and Zhiboedov derived a sequence of Ward identities, which enabled them to express all the planar $3$-point functions in terms of $\tilde{N}$, $\tlambda \propto \tilde{N}a_2$
and also in terms of $a_3$ in the quasi-boson case.

The $3$-point functions of single-trace primaries in the quasi-boson and quasi-fermion theories are constrained to contain at most three different conformal structures in the large $N$ limit \cite{Maldacena:2012sf},
\begin{align}
\langle J^{(s_1)} J^{(s_2)} J^{(s_3)} \rangle &= \alpha_{s_1 s_2 s_3}\langle J^{(s_1)} J^{(s_2)} J^{(s_3)} \rangle_{\text{bos.}} + \beta_{s_1 s_2 s_3}\langle J^{(s_1)} J^{(s_2)} J^{(s_3)} \rangle_{\text{fer.}} \notag\\
 &\quad + \gamma_{s_1 s_2 s_3}\langle J^{(s_1)} J^{(s_2)} J^{(s_3)} \rangle_{\text{odd}} \ed
 \label{3pnt}
\end{align}
In the above equation $\langle\cdot\rangle_{\text{bos.}}$ and $\langle\cdot\rangle_{\text{fer.}}$ refer to the correlators in the theory of a real free boson and fermion respectively, while $\langle\cdot\rangle_{\text{odd}}$ refers to independent conformally-invariant contributions to the correlators that appear only in interacting theories. Explicit expressions for these structures can be found in \cite{Giombi:2011rz,Costa:2011mg,Maldacena:2011jn}.

To extract meaningful information from the $3$-point functions \eqref{3pnt} one has to specify the precise normalization of the operators. In \cite{Maldacena:2012sf} a normalization was chosen such that the $2$-point functions of currents are equal in the quasi-boson and quasi-fermion theories,
\begin{align}
\langle J^{(s)}(x) J^{(s)}(0) \rangle = \tilde{N} \langle J^{(s)}(x) J^{(s)}(0) \rangle_{\text{bos.}} \ecq s \neq 0 \ed
\label{2pntCurrents}
\end{align}
The $2$-point functions of the scalar operators were chosen to be
\begin{align}
\langle J^{(0)}(x) J^{(0)}(0) \rangle &= \frac{\tilde{N}}{1+\tlambda^2} \langle J^{(0)}(x) J^{(0)}(0) \rangle_{\text{bos.}} \ec \label{2pntQB}\\
\langle \tilde{J}^{(0)}(x) \tilde{J}^{(0)}(0) \rangle &= \frac{\tilde{N}}{1+\tlambda_{\text{qf}}^2} \langle \tilde{J}^{(0)}(x) \tilde{J}^{(0)}(0) \rangle_{\text{fer.}} \ed \label{2pntQF}
\end{align}

With these normalizations the coefficients of the conformal structures in the $3$-point functions \eqref{3pnt} in the quasi-boson case were found to be (in the large $N$ limit)
\begin{align}
\alpha_{s_1 s_2 s_3} &= \tilde{N}\frac{1}{1+\tlambda^2} \ecq \!\!\beta_{s_1 s_2 s_3} = \tilde{N}\frac{\tlambda^2}{1+\tlambda^2} \ecq \gamma_{s_1 s_2 s_3} = \tilde{N}\frac{\tlambda}{1+\tlambda^2} \ec \notag \\
\alpha_{s_1 s_2 0}   &= \tilde{N}\frac{1}{1+\tlambda^2} \ecq \gamma_{s_1 s_2 0}  = \tilde{N}\frac{\tlambda}{1+\tlambda^2} \ec \label{3pntQB}\\
\alpha_{s_1 0 0}     &= \tilde{N}\frac{1}{1+\tlambda^2} \ecq \notag \\
\alpha_{0 0 0}       &= \tilde{N}\frac{1}{(1+\tlambda^2)^2} + z \left(\frac{\tilde{N}}{1+\tlambda^2}\right)^3 a_3 \ec \notag
\end{align}
where $z$ is a specific constant (denoted by $z_1 (n_0^{\text{free boson}})^3$ in \cite{Maldacena:2012sf}).

Similarly, for the quasi-fermion theories
\begin{align}
\alpha_{s_1 s_2 s_3} &= \tilde{N}\frac{\tlambda_{\text{qf}}^2}{1+\tlambda_{\text{qf}}^2} \ecq \!\!\beta_{s_1 s_2 s_3} = \tilde{N}\frac{1}{1+\tlambda_{\text{qf}}^2} \ecq \gamma_{s_1 s_2 s_3} = \tilde{N}\frac{\tlambda_{\text{qf}}}{1+\tlambda_{\text{qf}}^2} \ec \notag \\
\beta_{s_1 s_2 \tilde{0}}    &= \tilde{N}\frac{1}{1+\tlambda_{\text{qf}}^2} \ecq \gamma_{s_1 s_2 \tilde{0}}  = \tilde{N}\frac{\tlambda_{\text{qf}}}{1+\tlambda_{\text{qf}}^2} \ec \label{3pntQF}\\
\beta_{s_1 \tilde{0} \tilde{0}}      &= \tilde{N}\frac{1}{1+\tlambda_{\text{qf}}^2} \ec \notag \\
\gamma_{\tilde{0} \tilde{0} \tilde{0}}       &= 0 \ed \notag
\end{align}
The coefficients which are not specified in the above equations correspond to structures which are inconsistent with conformal invariance.

To obtain the correlators of the critical bosonic vector model with Chern-Simons interactions it is convenient to define $\tilde{J}^{(0)}_{\text{crit. bos.}} = \tlambda_{\text{qf}}\tilde{J}^{(0)}$. In terms of $\tilde{J}^{(0)}_{\text{crit. bos.}}$ the $2$-point function \eqref{2pntQF} and $\beta_{s_1 \tilde{0} \tilde{0}}$ are multiplied by $\tlambda_{\text{qf.}}^2$, while $\beta_{s_1 s_2 \tilde{0}}$ and $\gamma_{s_1 s_2 \tilde{0}}$ are multiplied by $\tlambda_{\text{qf}}$. The correlators of the critical model with $\lambda=0$ are then obtained by taking the $\tlambda_{\text{qf}} \to \infty$ limit in \eqref{2pntQF} and \eqref{3pntQF}. The correlators of the critical fermionic vector model are obtained from \eqref{2pntQB} and \eqref{3pntQB} in the same way.

The normalization of $J^{(0)}$ we use in this paper \eqref{currents} is different than the one in \eqref{2pntQB}, and the precise relation between the two normalizations will be fixed in section \ref{micro}. Note that in \cite{Maldacena:2012sf} only correlation functions of even-spin operators (that appear both in the $O(N)$ and in the $U(N)$ theory) were computed, while many of our results below involve the spin-one current $J^{(1)}$. However, since the odd-spin currents are part of the same high-spin symmetry algebra as the even-spin currents, and the results of \cite{Maldacena:2012sf} \eqref{2pntCurrents}, \eqref{3pntQB}, \eqref{3pntQF} are (as far as the coupling-dependence goes) independent of the spins for all $s > 0$, we assume that the same results hold also for odd-spin correlation functions. Our results will provide a consistency check on this natural assumption.

\section{Exact Scalar Correlators}
\label{scalars}

In this section we compute 2-point and 4-point scalar correlators exactly in planar perturbation theory. These correlation functions are not gauge-invariant, but they will be useful for computing
gauge-invariant correlation functions in the next section.
The computations are shown in some detail, to illustrate the techniques that will be useful later on.

\subsection{Scalar Propagator}
\label{prop}

\begin{figure}[!ht]
\centering
  \includegraphics[width=1.1\textwidth]{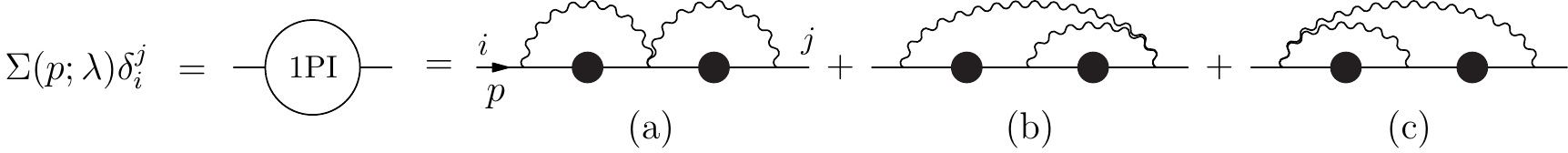}
  \caption{Bootstrap equation for the scalar self-energy. A filled circle denotes the full scalar propagator.}
  \label{fig:BS}
\end{figure}

Let us denote the self-energy of the scalar field by $\Sigma(p;\lambda)$, such that the full propagator is
\begin{align}
  \langle \phi^\dagger_i(p) \phi^j(q) \rangle &=
  \frac{\delta_i^j}{p^2 - \Sigma(p;\lambda)}
  \cdot (2\pi)^3 \delta^3(p+q) \ed
\end{align}
We will compute the self-energy in light-cone gauge at large $N$ by solving its bootstrap equation, shown in figure \ref{fig:BS}. Only non-vanishing planar diagrams are shown. The diagram of order $\lambda$, with a single gluon line, vanishes due to parity\footnote{While the Chern-Simons interactions break parity, our theory is invariant under the combined operation of parity plus $\lambda \mapsto -\lambda$. Corrections to the self-energy that come with an odd power of $\lambda$ must therefore be accompanied by an $\epsilon_{\mu\nu\rho}$ symbol, but in $\Sigma(p;\lambda)$ there is only a single external momentum to saturate it.}. Diagrams that involve the $A\wedge A \wedge A$ vertex vanish in light-cone gauge. Diagrams in which two seagull vertices are connected by a gluon line also vanish; this is because the seagull term $\phi^\dagger A^\mu A_\mu \phi$ reduces in this gauge to $\phi^\dagger A_3 A_3 \phi$, and the only non-vanishing gluon propagator is $\langle A_3 A_+\rangle $. All other diagrams which do not appear in figure \ref{fig:BS}, including those involving a $(|\phi|^2)^3$ vertex, are subleading in $1/N$.

Let us solve the equation of figure \ref{fig:BS}. The diagrams on the right-hand side of the bootstrap equation are given by
\begin{align}
  \mathrm{(a)} &= - (T^a\{T^a,T^b\}T^b)_{ji}
  \int \frac{d^dk}{(2\pi)^d}\frac{d^dl}{(2\pi)^d}
  \frac{(l+p)^{\mu}G_{\mu\nu}(l-p)G^{\nu}_{~\rho}(k-p)(k+p)^{\rho}}
  {[k^2-\Sigma(k)][l^2-\Sigma(l)]} \ec\notag\\
  \mathrm{(b)}+\mathrm{(c)} &= 2 (\{T^a,T^b\}T^aT^b)_{ji}
  \int \frac{d^dk}{(2\pi)^d}\frac{d^dl}{(2\pi)^d}
  \frac{(l+p)^{\mu}G_{\mu\nu}(l-p)G^{\nu}_{~\rho}(k-l)(k+l)^{\rho}}
  {[k^2-\Sigma(k)][l^2-\Sigma(l)]} \ed
\end{align}
Using the gluon propagator \eqref{gluon-prop}, in the planar limit we have
\begin{align}
  \mathrm{(a)} &= - 4(\pi\lambda)^2 \delta^j_i
  \int \frac{d^dk}{(2\pi)^d}\frac{d^dl}{(2\pi)^d}
  \frac{(l+p)^+(k+p)^+}{(l-p)^+(k-p)^+}
  \frac{1}{[k^2-\Sigma(k)][l^2-\Sigma(l)]} \ec \notag\\
  \mathrm{(b)}+\mathrm{(c)} &= 8 (\pi\lambda)^2 \delta^j_i
  \int \frac{d^dk}{(2\pi)^d}\frac{d^dl}{(2\pi)^d}
  \frac{(l+p)^+(k+l)^+}{(l-p)^+(k-l)^+}
  \frac{1}{[k^2-\Sigma(k)][l^2-\Sigma(l)]} \ed
  \label{abc}
\end{align}
Notice that light-cone gauge preserves the symmetry of rotations in the 1-2 plane, which act as $A_- \mapsto e^{i\theta} A_-$. Hence, $\Sigma(p)=\Sigma(p_s,p_3)$, where $p_s = \sqrt{p_1^2 + p_2^2}$. Further, the expressions \eqref{abc} are independent of $p_3$, and therefore we can guess that $\Sigma(p;\lambda) = f(\lambda) p_s^2$.

Let us now focus on diagram (a). The integrals over the $p_3$ component of the loop momenta in \eqref{abc} can be carried out,
\begin{align}\label{eq:k3Integral}
  \int \frac{d^{1-\epsilon}k_3}{(2\pi)^{1-\epsilon}}
  \frac{1}{k^2 - \Sigma(k)} =
  \int \frac{d^{1-\epsilon}k_3}{(2\pi)^{1-\epsilon}}
  \frac{1}{(k_3)^2+(1-f(\lambda))k_s^2} \;&\longrightarrow\;
  \frac{1}{2 k_s \sqrt{1-f(\lambda)}} \ec
\end{align}
where we have taken $\epsilon \to 0$ at the end. For the remaining integrals in the 1-2 momentum plane we use polar coordinates, with $k^\pm = k_s e^{\pm i \theta_k} / \sqrt{2}$. The angular integral can be treated as a contour integral, for example
\begin{align}
  \int_0^{2\pi} d\theta_k \frac{k^+ + p^+}{k^+ - p^+} =
  \oint \frac{dz}{iz}
  \frac{z + \sqrt{2}p^+/k_s}{z - \sqrt{2}p^+/k_s} =
  2\pi[2\Theta(k_s-p_s)-1] \ec
\end{align}
where $z = e^{i\theta_k}$ is integrated over the unit circle and $\Theta$ is the step function. The remaining radial integrals (performed up to the cutoff $\Lambda$) are trivial, and the result is
\begin{align}
  \mathrm{(a)} &= -\frac{\lambda^2 \delta^j_i}{4(1-f(\lambda))}
  (2 p_s - \Lambda)^2 \ecq
  \mathrm{(b)}+\mathrm{(c)} = \frac{\lambda^2 \delta^j_i}{1-f(\lambda)}
  p_s (p_s - \Lambda) \ed
\end{align}
The self-energy $(a)+(b)+(c)$ is therefore a pure divergence,
\begin{align}
  \Sigma &= -\frac{\lambda^2}{4(1-f(\lambda))} \Lambda^2 \ed
\end{align}
This divergence is subtracted with a mass counterterm $\phi^\dagger\phi$, which is determined uniquely by conformal invariance. In the continuum theory the result is therefore $\Sigma = 0$, namely the scalar propagator in our gauge does not receive any corrections.

\subsection{Scalar 4-Point Function}

\begin{figure}
  \centering
  \includegraphics[width=0.5\textwidth]{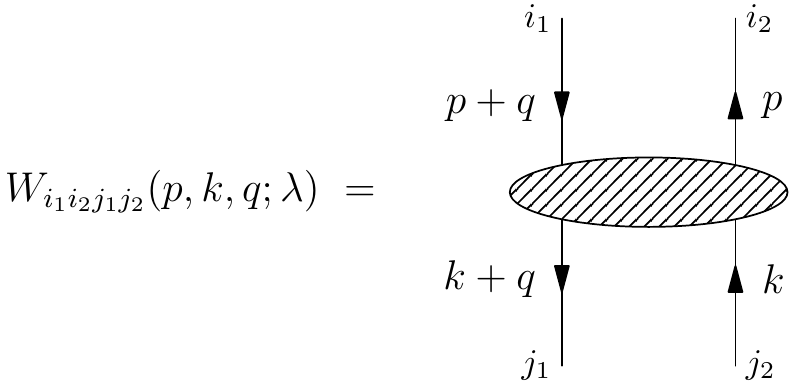}
  \caption{Connected diagrams in scalar 4-point function.}
  \label{fig:W}
\end{figure}

In this section we compute the connected scalar 4-point function
\begin{equation} \label{fourpt}
\langle \phi^{i_1}(p+q) \phi^{\dagger}_{i_2}(-p) \phi^{\dagger}_{j_1}(-k-q) \phi^{j_2}(k) \rangle,
\end{equation}
shown in figure \ref{fig:W}.
From now on the overall factor $(2\pi)^3 \delta^3(\sum k)$ is implicit in all correlators we write.
Without loss of generality we focus on the terms proportional to $\delta_{i_1 i_2} \delta_{j_1 j_2}$ (the other terms are related to this by permutations of the momenta).

\begin{figure}
  \centering
  \includegraphics[width=0.5\textwidth]{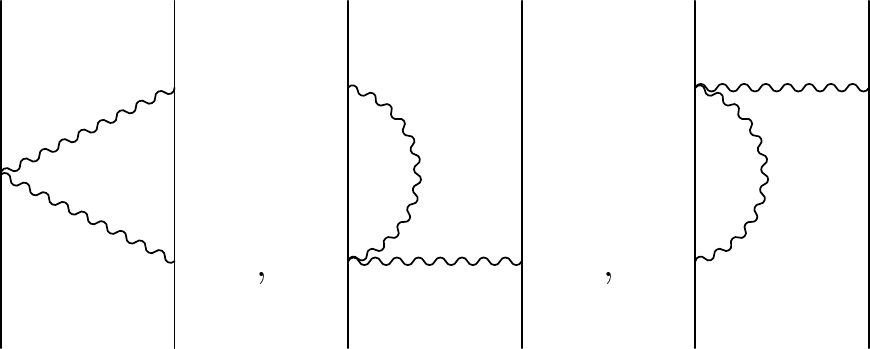}
  \caption{1-loop diagrams that include the $A^2\phi^2$ vertex, up to reflections.}
  \label{fig:WA}
\end{figure}

\begin{figure}
\centering
  \includegraphics[width=0.8\textwidth]{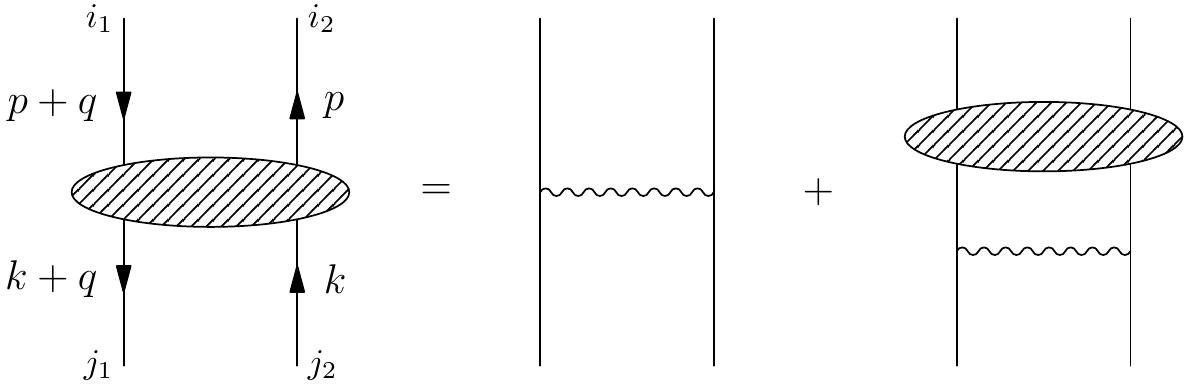}
  \caption{Bootstrap equation for the connected 4-point function, when $q^\pm=0$.}
  \label{fig:Wbs}
\end{figure}

Let us first consider the sum of 1-loop diagrams that contribute to $W$ (the correlator \eqref{fourpt} without the delta function) and
that include an $A^2\phi^2$ vertex. There are six such diagrams: the ones shown in
figure \ref{fig:WA}, and their reflections along the vertical axis.
It is easy to compute these diagrams using the methods of the previous section; their sum is
\begin{align}
  2\pi N \lambda^2 \left\{
    \frac{2 q^+}{(k-p)^+} \left[
        (k+q)_s - (p+q)_s - k_s + p_s
    \right]
    - \Lambda
  \right\} \ed
\end{align}
The linear divergence is subtracted by a $(\phi^\dagger\phi)^2$ counterterm, which is uniquely determined by conformal invariance. From now on we will take $q^\pm=0$ for simplicity, even though this is not the most general 4-point function. For this choice of momentum the seagull vertex does not contribute to the 4-point function, and $W$ becomes a sum of ladder diagrams. The bootstrap equation for the 4-point function with $q^+=0$ is shown in figure \ref{fig:Wbs}.
In writing it we used the fact that the scalar propagator does not receive quantum corrections in our gauge.

Let us now restrict the form of $W_{i_1 i_2 j_1 j_2}(p,k,q;\lambda)$.
Using dimensional analysis we can write
\begin{align}
  W_{i_1 i_2 j_1 j_2}(p,k,q;\lambda) =
  \delta_{i_1 i_2}\delta_{j_1 j_2} \, |q_3| \,
  \tilde{W}\!\left( \frac{k_s}{q_3}, \frac{p_s}{q_3}, \dots ; \lambda \right) \ed
  \label{Wdef}
\end{align}
By computing the first few contributions to $\tilde{W}$ explicitly, we can see that they depend only on the variables
\begin{align} \label{xyzdef}
 x=\frac{k_s}{|q_3|} \ecq
 y=\frac{p_s}{|q_3|} \ecq
 z=\frac{(k+p)^+}{(k-p)^+} \ecq
 \Lambda'=\frac{\Lambda}{|q_3|} \ecq
 \hlambda = \lambda\ {\mathrm{sign}}(q_3)
 \ec
\end{align}
and that the dependence on $z$ is at most linear\footnote{We choose to work with parity-invariant variables. Up to order $\lambda^3$, we find
\begin{align*}
  \tilde{W}|_{o(\lambda)} &= -4\pi i \frac{\hlambda}{N} z \ec \notag \\
  \tilde{W}|_{o(\lambda^2)} &= 4\pi \frac{\hlambda^2}{N} \left[
        \arctan(2\Lambda') +
        2 z \left(
            \arctan(2y) - \arctan(2x)
        \right)
    \right] \ec \notag \\
  \tilde{W}|_{o(\lambda^3)} &= 8\pi i \frac{\hlambda^3}{N} \Big[
        \arctan(2\Lambda') \left( \arctan(2y) - \arctan(2x) \right) +
        z
        \left( \arctan(2y) - \arctan(2x) \right)^2
    \Big] \ec
\end{align*}
where $x,y,z$ are defined above.
}.
Let us assume that this holds to all orders, and write
\begin{align}
  \tilde{W}(x,y,z;\hlambda) = \tilde{W}_0(x,y;\hlambda)
  + z \,\tilde{W}_1(x,y;\hlambda) \ed
\end{align}
Returning to the bootstrap equation, after carrying out the loop integrals over the 3-component and the angular directions, the equation can be written as
\begin{align}
  \tilde{W}_0 + \tilde{W}_1 z &=
  -4\pi i \frac{\hlambda}{N} z + 2 i \hlambda (I_0 + I_1) \ec \notag \\
  I_0 &= |q_3| \left[ 2 \int_{k_s}^\Lambda - \int_0^\Lambda \right] dk_s' \,
  \frac{\tilde{W}_0 \left( x', y \right)}{4k_s'^2 + q_3^2} \ec \notag \\
  I_1 &= |q_3| \left[ \int_0^\Lambda + 2 z \int_{k_s}^{p_s} \right] dk_s' \,
  \frac{\tilde{W}_1(x',y)}{4k_s'^2 + q_3^2}
  \ec
  \label{Wbs}
\end{align}
where $x' = \frac{k_s'}{|q_3|}$. Let us first solve for $\tilde{W}_1$ by equating the coefficients of $z$ and differentiating with respect to $k_s$. The result is
\begin{align}
  \tilde{W}_1(x,y) &= C_1(y) e^{-2i\hlambda \arctan(2x)} \ed
\end{align}
Here $C_1(y)$ is an integration ``constant'' that is determined, by plugging $\tilde{W}_1$ back into \eqref{Wbs}, to be
\begin{align}
  C_1(y) &= -\frac{4\pi i \hlambda}{N}
  e^{2i \hlambda \arctan(2y)} \ed
\end{align}
One can now solve for $\tilde{W}_0$ similarly, and plug the result in \eqref{Wdef}. We find
\begin{align}
  \left. W_{i_1 i_2 j_1 j_2}(|q|,x,y,z;\hlambda) \right|_{q^\pm=0} &=
  \delta_{i_1 i_2} \delta_{j_1 j_2} \frac{4\pi\hlambda |q_3|}{N}
  \left[
  \tan \left( \hlambda \arctan\left( 2\Lambda' \right) \right) - i z
  \right]
  \times \notag \\ &\quad
  \exp \! \left[
  2i \hlambda \left( \arctan(2y) - \arctan(2x) \right)
  \right] \ed
  \label{W}
\end{align}

\section{Exact Gauge-Invariant Correlators}
\label{correlators}

In this section we compute several exact 2-point and 3-point functions of single-trace primary operators in momentum space. Using the results of section \ref{scalars} one can compute any such correlator, with all external momenta pointing in the 3-direction, by computing a finite number of integrals. A useful step is to first compute exact vertices of the form $\langle J^{(s)}(-q)\phi\phi^\dagger\rangle$, since they encode all the information of the scalar 4-point function inside 2-point and 3-point gauge-invariant correlators. We begin by computing correlators of $J^{(0)} \equiv \phi^\dagger \phi$, and then generalize to other operators.

\subsection{$\left< J^{(0)} J^{(0)} \right>$}

Let us compute the $J^{(0)}$ vertex
\begin{align}
  \langle J^{(0)}(-q) \phi^\dagger_j(k) \phi^i \rangle &=
  \delta^i_j V_0(q,k;\lambda)
  \ecq
\end{align}
for the special case of $q^{\pm}=0$, where we can use the results of the previous section. (From here
on we do not always explicitly write the momentum of the last operator in our correlation functions, which is fixed by momentum conservation.)
\begin{figure}
  \centering
  \includegraphics[width=0.7\textwidth]{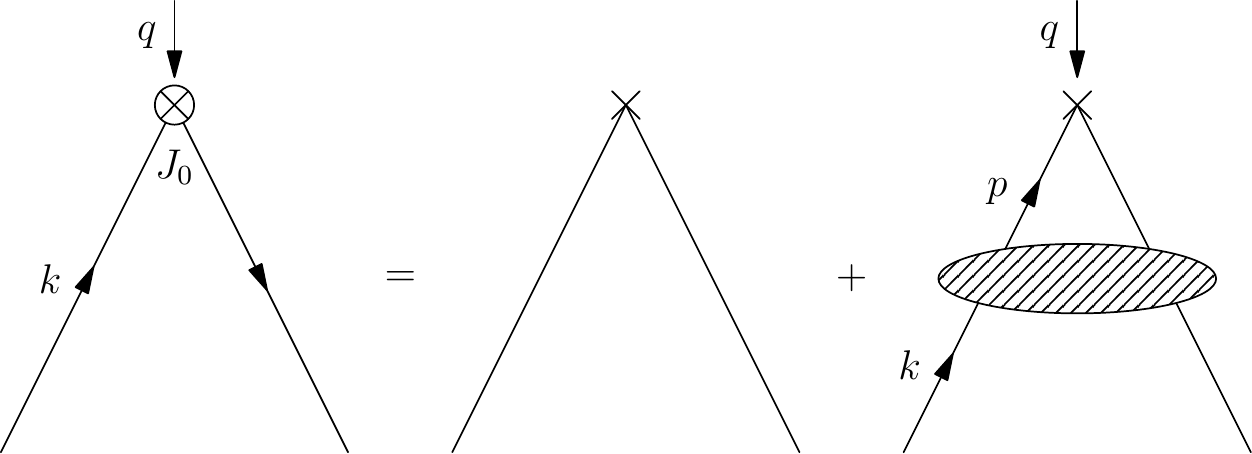}
  \caption{The vertex $\langle J^{(0)} \phi \phi^\dagger\rangle$. A cross denotes a $J^{(0)}$ insertion in the free theory, and a circled cross denotes the exact vertex. The hatched ellipse denotes diagrams in which the 4 scalar lines are connected.}
  \label{fig:J0vertex}
\end{figure}
As shown in figure \ref{fig:J0vertex}, we can write this vertex as a sum of a free piece, and a piece from the connected 4-scalar function $W$. The free piece is just $\delta^i_j$. The connected piece is given by (using the variables of \eqref{xyzdef})
\begin{align}
  N |q_3| \delta^i_j
  \int \frac{d^dp}{(2\pi)^d} \frac{\tilde{W}(x,y,z;\hlambda)}{p^2 (p+q)^2}
  = \delta^i_j \left[\frac{2 e^{-2i\hlambda \atan(2x)}}{1 + e^{-2i\hlambda \atan(2\Lambda')}} -1 \right] \ec
  \label{J0vertCon}
\end{align}
where the integral can be computed by the same method as before. Crucially, the radial integration can be carried out analytically. Adding the free piece, we find
\begin{align}
  V_0(x;\hlambda) =
  \frac{2 e^{-2i\hlambda \atan(2x)}}{1 + e^{-2i\hlambda \atan(2\Lambda')}} \ed
  \label{V0}
\end{align}

\begin{figure}
  \centering
  \includegraphics[width=0.3\textwidth]{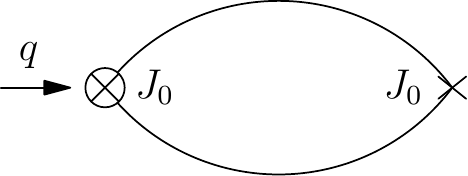}
  \caption{The diagrams contributing to $\langle J^{(0)}(-q) J^{(0)}\rangle$.}
  \label{fig:J0J0}
\end{figure}

It is now easy to compute the $J^{(0)}$ 2-point function, shown in figure \ref{fig:J0J0}. Note that replacing a single $J^{(0)}$ insertion by the exact vertex \eqref{V0} accounts for all the diagrams without any double-counting. Thus,
\begin{equation}
  \langle J^{(0)}(-q) J^{(0)} \rangle =
  N \int \frac{d^dk}{(2\pi)^d} \frac{V_0(x;\hlambda)}{k^2 (k+q)^2} \to \frac{N}{4}\frac{1}{|q_3|}
  \frac{\tan\!\big(\frac{\pi\hlambda}{2}\big)}{\pi\hlambda} \ec \label{eq:J0J0}
\end{equation}
where in writing the final result we took $\Lambda \to \infty$. Note that the result is even under
$\lambda \to -\lambda$, as expected. The obvious Lorentz-invariant generalization of this result to any $q$ is
\begin{equation}
  \langle J^{(0)}(-q) J^{(0)} \rangle =
  \frac{N}{4}\frac{1}{|q|}
  \frac{\tan\!\big(\frac{\pi\lambda}{2}\big)}{\pi\lambda} \ed \label{eq:J0J0n}
\end{equation}

The momentum dependence of \eqref{eq:J0J0n} is determined by conformal invariance; in position space this translates into $\langle J^{(0)}(x) J^{(0)}(y) \rangle \propto 1 / (x - y)^2$. The dependence on $\lambda$ is not fixed by the symmetries, and we observe that the 2-point function diverges at $\lambda=1$, and becomes negative afterwards, suggesting that our theories only make sense up to $\lambda=1$. As mentioned above, if we define the Chern-Simons level using a regularization involving a Yang-Mills theory, the value $\lambda=1$ is actually the maximal allowed value, so this is not too surprising (similar results were found also for fermions coupled to Chern-Simons theory in \cite{Giombi:2011kc}).

\subsection{$\left< J^{(0)} J^{(0)} J^{(0)} \right>$}

\begin{figure}
  \centering
  \includegraphics[width=1\textwidth]{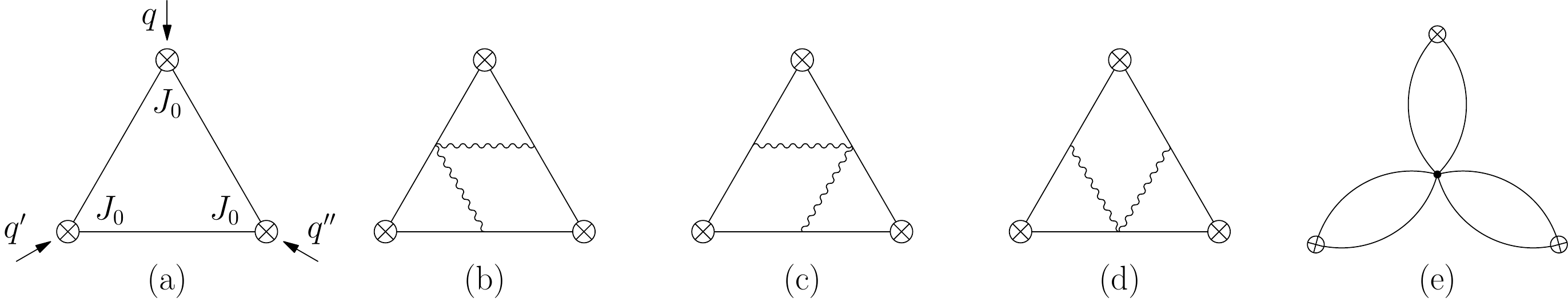}
  \caption{The diagrams contributing to $\langle J^{(0)}(-q) J^{(0)}(-q') J^{(0)}(-q'') \rangle$.}
  \label{fig:J0J0J0}
\end{figure}

Using the exact $J^{(0)}$ vertex \eqref{V0} we can compute the 3-point function of scalar operators $\langle J^{(0)}(-q) J^{(0)}(-q') J^{(0)}(-q'') \rangle$, with all momenta $q,q',q''$ in the $x^3$-direction. Using the vertices defined above, all contributions to this 3-point function are included in the diagrams of figure \ref{fig:J0J0J0}. All the diagrams turn out to be finite, so we remove the cutoff in the expressions below. Diagram $(a)$ evaluates to
\begin{align}
  \mathrm{(a)} &= 2N \int \frac{d^3k}{(2\pi)^3} \frac{V_0(q,k) V_0(q',k+q') V_0(q'',k-q'')}{k^2(k+q)^2(k-q'')^2} \notag\\
    &= \frac{N \tan\left(\frac{\pi\lambda}{2}\right)}{2\pi\lambda\cos^2\left(\frac{\pi\lambda}{2}\right)} \frac{1}{|q||q'||q''|} \ed
\end{align}

Summing over diagrams (b), (c) and (d), and integrating over the 3 and angular directions gives
\begin{align}
  \mathrm{(b)+(c)+(d)} &= -\frac{\lambda^2 N}{\pi} \int_0^{\infty}dk_s dl_s dp_s \frac{V_0(q,k)V_0(q',l)V_0(q'',p)}{(4k_s^2+q^2)(4l_s^2+q'^2)(4p_s^2+q''^2)} \notag \\
            &~~~ \big\{ \left(2\Theta(k_s-p_s)-1\right)\left(2\Theta(l_s-p_s)-1\right) \notag\\
            &\quad + \left(2\Theta(p_s-k_s)-1\right)\left(2\Theta(l_s-k_s)-1\right) \notag\\
            &\quad + \left(2\Theta(k_s-l_s)-1\right)\left(2\Theta(p_s-l_s)-1\right) \big\} \notag \\
            &= -\frac{\lambda^2 N}{\pi} \left(\int_0^{\infty}dk_s \frac{V_0(q,k)}{4k_s^2+q^2}\right) \left(\int_0^{\infty}dl_s \frac{V_0(q',l)}{4l_s^2+q'^2}\right) \left(\int_0^{\infty}dp_s \frac{V_0(q'',p)}{4p_s^2+q''^2}\right) \notag \\
            &= -\frac{N\tan^3\left(\frac{\pi\lambda}{2}\right)}{8\pi\lambda} \frac{1}{|q||q'||q''|} \ec
\end{align}
where in the second equality we used the fact that
\begin{align}
    \left(2\Theta(k_s-p_s)-1\right)\left(2\Theta(l_s-p_s)-1\right) &\quad \notag\\
    + \left(2\Theta(p_s-k_s)-1\right)\left(2\Theta(l_s-k_s)-1\right) &\quad \notag\\
    + \left(2\Theta(k_s-l_s)-1\right)\left(2\Theta(p_s-l_s)-1\right) &= 1 \ed
\end{align}

The remaining diagram $\mathrm{(e)}$ contributes if $\lambda_6 \neq 0$.
It gives
\begin{align}
  \mathrm{(e)} &= -\frac{N \lambda_6 \tan^3\left(\frac{\pi\lambda}{2}\right)}{64\pi^3\lambda^3} \frac{1}{|q||q'||q''|} \ed
\end{align}
Note that $\langle J^{(0)} J^{(0)} J^{(0)} \rangle$ is the only 3-point function which receives a contribution from the $(\phi^{\dag}\phi)^3$ vertex in the planar limit. This is since diagram $\mathrm{(e)}$, when computed in momentum space, factorizes into a product of three 2-point functions of $J^{(0)}$ with the inserted operators: $\langle J^{(s)} J^{(s')} J^{(s'')} \rangle_{\mathrm{(e)}} \propto \langle J^{(s)} J^{(0)} \rangle \langle J^{(s')} J^{(0)} \rangle \langle J^{(s'')} J^{(0)} \rangle$, and these 2-point functions vanish by conformal invariance unless $s=s'=s''=0$.

Summing up all the contributions we obtain
\begin{align}
\langle J^{(0)}(-q) J^{(0)}(-q') J^{(0)}(-q'') \rangle = \frac{N}{2\pi\lambda}\left[ \frac{\tan\left(\frac{\pi\lambda}{2}\right)}{\cos^2\left(\frac{\pi\lambda}{2}\right)} - \frac{1}{4}\tan^3\left(\frac{\pi\lambda}{2}\right)\left(1+\frac{\lambda_6}{8\pi^2\lambda^2}\right) \right] \frac{1}{|q||q'||q''|} \ec \label{eq:J0J0J0}
\end{align}
which has the correct momentum dependence required by conformal invariance (it is uniquely determined by the result we computed for $q^{\pm}=q'^{\pm}=q''^{\pm}=0$).

\subsection{$\left< J^{(1)} J^{(1)} \right>$}
\label{sec:J1J1}

Next, we compute two correlators that involve $J^{(1)}$.

\begin{figure}
  \centering
  \includegraphics[width=0.3\textwidth]{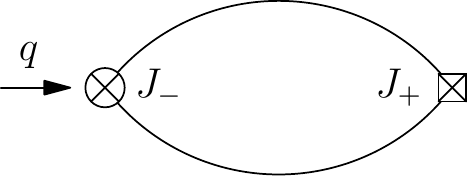}
  \caption{The diagrams contributing to $\langle J_-(-q) J_+ \rangle$.}
  \label{fig:JmJp}
\end{figure}

\begin{figure}
  \centering
  \includegraphics[width=0.9\textwidth]{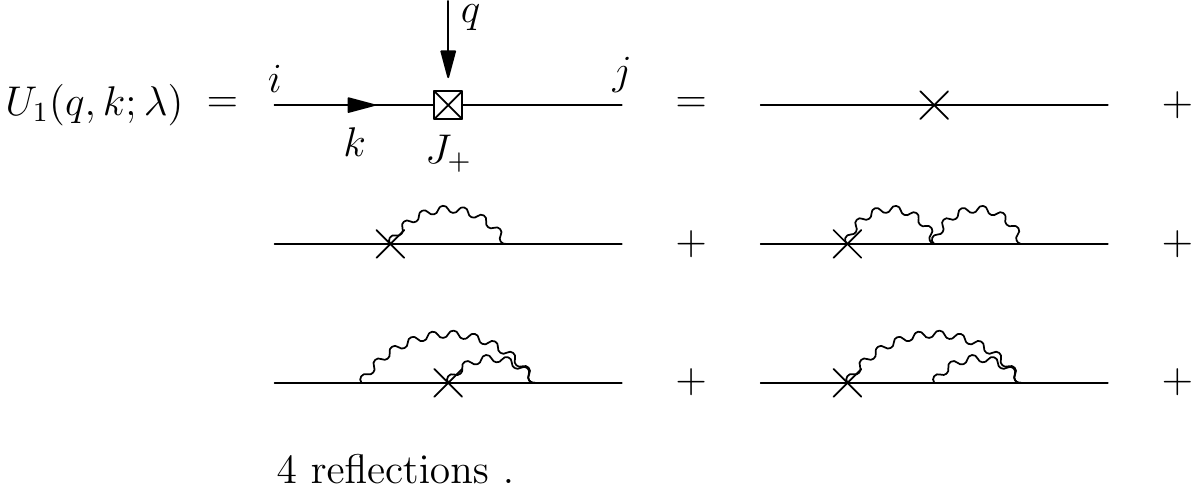}
  \caption{Diagrams of $\langle J_+ \phi^\dagger \phi \rangle$ in which a gauge
  field is connected to $J_+$.}
  \label{fig:Jp}
\end{figure}

In this subsection we compute $\langle J_-(-q) J_+ \rangle$, with $q^\pm=0$\footnote{Note that when $q^\pm=0$, current conservation implies $q_3 \langle J_3(q) \cdots \rangle = 0$ up to contact terms, and therefore correlators involving the $J_3$ component of $J^{(1)}$ do not contain information that is useful for our purposes.}. For
this purpose let us first compute the $J_-$ vertex, namely $\langle J_-(-q) \phi^\dagger \phi \rangle$ with $q^\pm=0$. In fact, it is no more work to compute the $J_{-\cdots -}$ vertex
\begin{align}
  \langle J_{-\cdots -}(-q) \phi^\dagger_j(k) \phi^i \rangle &=
  \delta^i_j V_s(q,k;\lambda) \ec
  \label{Jsvert}
\end{align}
for any current $J^{(s)}$ with $s>0$, as long as $q^\pm=0$. Indeed, in our gauge the bootstrap equation for this has the same form as the equation for the $J^{(0)}$ vertex (figure \ref{fig:J0vertex}), since $A_-=0$ so there are no additional diagrams where a gauge field is connected to $J^{(s)}$. Further, by Lorentz invariance the free piece in \eqref{Jsvert} is given (when $q^\pm=0$) by $\alpha_s (k^+)^s$, where the factor $\alpha_s$ depends on the normalization of the current. In computing the connected piece of the bootstrap equation, the only modification to \eqref{J0vertCon} is to insert $\alpha_s (p^+)^s$ in the integral. The computation carries through, and we find for all $s \geq 1$
\begin{align}
  V_s &= \alpha_s (k^+)^s e^{2i\hlambda (\atan(2\Lambda') - \atan(2x))} \ed
  \label{Vs}
\end{align}
Note that $\alpha_1=2$ for the canonically-normalized $J^{(1)}$,

The 2-point function $\langle J_-(-q) J_+ \rangle$ receives the contributions shown in figure \ref{fig:JmJp}, where the circled insertion is the exact vertex given by $V_1$ of \eqref{Vs}, and the squared insertion is defined in figure \ref{fig:Jp}: it accounts for the diagrams in which a gauge field $A_+$ is connected to $J_+$. The squared insertion of figure \ref{fig:Jp} evaluates to
\begin{align}
  U_1(q_3,k;\lambda) &= \frac{k_s^2 + i \hlambda |q_3| k_s}{k^+} \delta^j_i \ed
  \label{U1}
\end{align}
The exact 2-point function $\langle J_-(-q) J_+\rangle $ is then given by
\begin{align}
  \langle J_-(-q) J_+\rangle  &=
  N \int \! \frac{d^dp}{(2\pi)^d} \, \frac{V_1(q,p) U_1(-q,p+q)}{p^2 (p+q)^2}
  \notag\\
  &\to \frac{i N |q_3|}{16}
  \frac{ e^{\pi i \hlambda} - 1 }{\pi \hlambda}
  + \frac{N}{4\pi} \Lambda \ed
  \label{JmJpraw}
\end{align}
In the second line we took $\Lambda \to \infty$, carefully keeping divergent terms.


This result has two peculiar features. First, the linear divergence is a contact term that violates conformal invariance. If we denote the background gauge field that couples to $J_\mu$ by $\cA_\mu$, then this divergence can be subtracted by a mass counterterm $\cA_\mu \cA^\mu$.
The second peculiar feature of \eqref{JmJpraw} is the appearance of a $\lambda$-odd part, equal to
\begin{align}
  \frac{i N q_3}{16} \frac{\cos(\pi\lambda) - 1}{\pi\lambda} \ed
\end{align}
This piece violates parity, while $\langle J J \rangle$ is parity-even in a conformal theory \cite{Osborn:1993cr,Giombi:2011rz}. However, this piece can come from a contact term $\langle J_\mu(q) J_\nu \rangle \sim \epsilon_{\mu\nu\rho} q^\rho$ that is conformally-invariant, and gives a derivative of a delta function in position space. It corresponds to the appearance of a Chern-Simons term $\frac{i\kappa}{4\pi} \int
\cA \wedge d\cA$ in the generating functional $F[\cA,\dots]$. In many cases, contact terms are scheme-dependent and therefore do not contain physical information. This is equivalent to saying that their values can be shifted arbitrarily by adding an appropriate counter-term. In our case this would correspond to shifting $\kappa$. However, since this is a Chern-Simons term we only have the freedom to shift it by an integer amount. The fractional part of this term (in units of $\kappa$) is therefore a physical observable \cite{Closset:2012vp}. It would be interesting to understand if this observable is constrained by the high-spin symmetry, like the other correlators discussed in \cite{Maldacena:2012sf}.

Removing both contact terms, we are left with the parity-even result
\begin{align}
  \langle J_-(-q) J_+\rangle &=
  - \frac{N|q|}{16} \frac{\sin(\pi \lambda)}{\pi \lambda} \ed
\end{align}
Note that this also changes sign at $\lambda=1$, consistent with our theory stopping to make sense (at least as a unitary theory) there.

\subsection{$\left< J^{(0)} J^{(1)} J^{(1)}\right>$}

\begin{figure}
  \centering
  \includegraphics[width=0.8\textwidth]{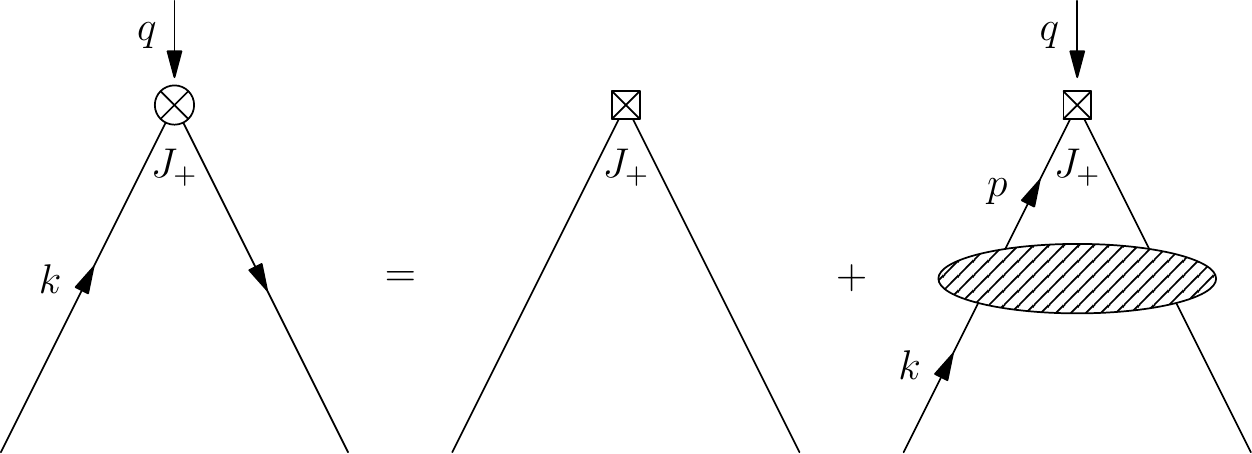}
  \caption{The vertex $\langle J_+(-q) \phi^\dagger(k) \phi \rangle$.}
  \label{fig:Jpvert}
\end{figure}

In this section we compute the 3-point function $\langle J^{(0)}(-q) J_+(-q') J_-(-q'')\rangle$, with $q,q',q''$ all in the 3-direction. We first need to compute the $J_+$ vertex,
\begin{align}
  \langle J_+(-q) \phi^\dagger_j(k) \phi^i \rangle &=
  \delta^i_j V_+(q,k;\lambda) \ec
  \label{Jpvert}
\end{align}
shown in figure \ref{fig:Jpvert}. It can be evaluated using \eqref{W} and \eqref{U1}, and the result is
\begin{align}
  V_+(q,k;\lambda) &=
  \frac{1}{4k^+} \left[4k_s^2 + q_3^2 - q_3^2
  e^{-2 i \hlambda \arctan\left( 2 x\right)} \right] \ed
\end{align}

\begin{figure}
  \centering
  \includegraphics[width=0.8\textwidth]{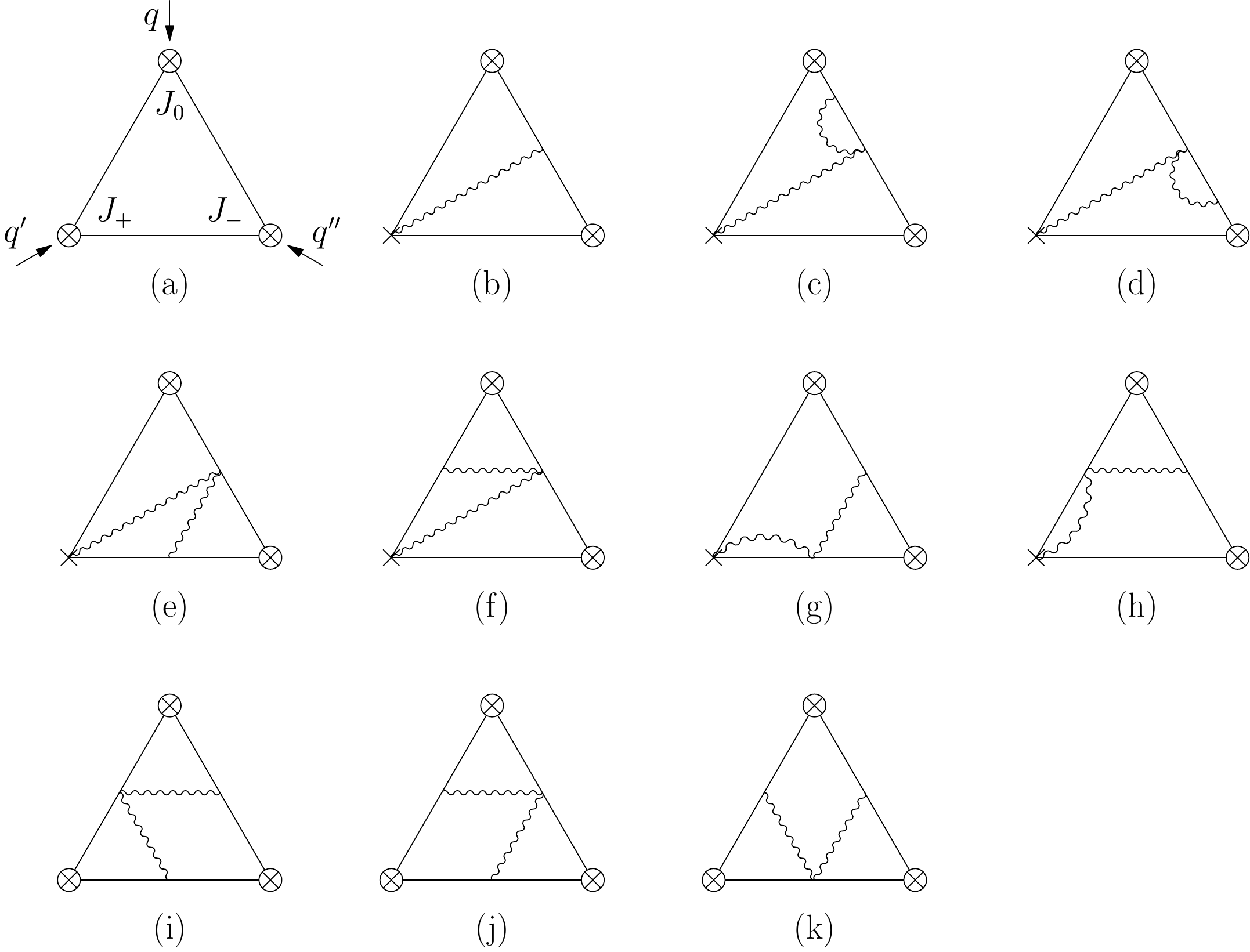}
  \caption{Diagrams contributing to $\langle J^{(0)}(-q) J_+(-q') J_-(-q'')\rangle$.}
  \label{fig:J1J1J0}
\end{figure}

Returning to the 3-point function, the diagrams which contribute to it are shown in figure \ref{fig:J1J1J0}.
In the diagrams that contain a seagull vertex, the final radial integrals cannot be performed analytically. However, using the step-function relation
\begin{align}
  \Theta(x-y)\Theta(y-z) + \Theta(y-x)\Theta(x-z) = \Theta(x-z)\Theta(y-z) \ec
\end{align}
one can see that these diagrams cancel in triples,
\begin{align}
  \mathrm{(c)}+\mathrm{(d)}+\mathrm{(e)}=
  \mathrm{(f)}+\mathrm{(g)}+\mathrm{(h)}=
  \mathrm{(i)}+\mathrm{(j)}+\mathrm{(k)}=0 \ed
\end{align}
The remaining diagrams $\mathrm{(a)},\mathrm{(b)}$ are given by
\begin{align}
  \mathrm{(a)} &= 2N \int \frac{d^3p}{(2\pi)^3}
  \frac{V_0(q,p-q) V_+(q',p) V_1(q'',p+q')}{p^2(p-q)^2(p+q')^2}
  \notag \\ &=
  \frac{N}{\pi} \frac{e^{\pi i \hlambda''}}{1+e^{-\pi i \hlambda}}
  \int_0^\Lambda \! dp_s \,
  e^{-2i\hlambda \arctan(2y)-2i\hlambda''\arctan(2y'')}
  \frac{12p_s^2 + q^2 + q'^2 + q q'}
  {(4p_s^2 + q^2)(4p_s^2 + q''^2)}
  \notag \\ &\quad +
  \frac{N}{\pi} \frac{e^{\pi i \hlambda''}}{1+e^{-\pi i \hlambda}}
  \frac{i q'
  \left[ 1 - e^{-\pi i ( \hlambda + \hlambda' + \hlambda'' )} \right]}
  {4 \lambda q (q + q')}
  \ec \notag \\
  \mathrm{(b)} &=
  8\pi i \lambda N \int \frac{d^3p}{(2\pi)^3} \frac{d^3k}{(2\pi)^3}
  \frac{V_0(q,p) V_1(q'',k-q'')}{p^2 (p+q)^2 k^2 (k-q'')^2} \frac{(p+k)_3}{(p-k)^+}
  \notag \\ &=
  \frac{N}{\pi} \frac{e^{\pi i \hlambda''}}{1+e^{-\pi i \hlambda}}
  \frac{q-q''}{q''}
  \int_0^\Lambda \! dp_s \,
  \frac{e^{-2i\hlambda \arctan(2y) - 2i\hlambda'' \arctan(2y'')}}{4p_s^2 + q^2}
  \notag \\ &\quad
  - \frac{N}{\pi} \frac{q-q''}{4\lambda q q''}
  \tan\left( \frac{\pi\hlambda}{2} \right)
  \ec
\end{align}
where now $\hlambda' = \lambda \, {\mathrm{sign}}(q'), x'=\frac{k_s}{|q'|}, y'=\frac{p_s}{|q'|}$, and similarly for $q''$.
In writing this we took $\Lambda \to \infty$ where possible to simplify the expressions. Since there are no divergences (as we shall see), this does not change the final result.

The remaining radial integrals cannot be computed separately, but the integral in the sum $\mathrm{(a)}+\mathrm{(b)}$ can be computed analytically. The final result is
\begin{align}
  \langle J^{(0)}(-q) J_+(-q') J_-(-q'')\rangle &=
  \frac{N}{\pi} \frac{e^{\pi i \hlambda''}}{1+e^{-\pi i \hlambda}}
  \frac
  {i q' \left[ 1 - e^{-\pi i ( \hlambda + \hlambda' + \hlambda'' )} \right]}
  {4 \lambda q (q + q')}
  - \frac{N}{\pi} \frac{q-q''}{4\lambda q q''}
  \tan\left( \frac{\pi\hlambda}{2} \right)
  \notag \\ &\quad
  + \frac{N}{\pi} \frac{e^{\pi i \hlambda''}}{1+e^{-\pi i \hlambda}}
  \frac{i(q+2q')}{4\lambda q'(q+q')}
  \left[ 1 - e^{-\pi i (\hlambda + \hlambda'')} \right] \ed
  \label{J0JpJm}
\end{align}
By conformal invariance, this correlator should have one parity-even and one parity-odd structure \cite{Giombi:2011rz}. Let us consider the $\lambda$-even and $\lambda$-odd parts separately.

\subsubsection{Even Structure}

Let $\chi \equiv \frac{\pi \lambda}{2}$. The $\lambda$-even part of
\eqref{J0JpJm} is
\begin{align}
  \langle \cdot \rangle_{\lambda\text{-even}} &=
  \frac{N}{8 \pi \lambda q q' (q+q')} \frac{1}{\cos({\mathrm{sign}}(q) \chi)}\Big\{
  (-q^2 + 2qq' + 2q'^2) \sin\! \big(\! {\mathrm{sign}}(q) \chi \big)
  \notag \\ &\quad
  - q'^2 \sin\! \big(\! {\mathrm{sign}}(q)\chi + 2 {\mathrm{sign}}(q')\chi \big)
  - q''^2 \sin\! \big(\! \mathrm{sign}(q)\chi + 2 {\mathrm{sign}}(q'')\chi \big)
  \Big\} \ed
\end{align}
By simple trigonometry we can write this as
\begin{align}
  \frac{N}{8} \frac{\sin(\pi\lambda)}{\pi\lambda} \left[
  \frac{|q|}{q'q''} + \frac{|q'|}{qq''} + \frac{|q''|}{qq'}
  \right]
  + \frac{N}{2}
  \frac{\sin^2\!\left( \frac{\pi\lambda}{2} \right)
  \tan\!\left( \frac{\pi\lambda}{2} \right)}
  {\pi\lambda}
  \frac{1}{|q|} \ed
  \label{J0JpJmEven}
\end{align}
By conformal invariance we expect only a single parity-even structure, so we expect that the second piece is a contact term. Indeed, it is easy to check that
\begin{align}
  \langle J^{(0)}(-q) J_\mu J_\nu\rangle &\sim
  \frac{\delta_{\mu\nu}}{|\vec{q}|}
  \label{cont}
\end{align}
is a conformally-invariant contact term (independent of the second momentum).

\subsubsection{Odd Structure}

Using similar methods, the odd part of \eqref{J0JpJm} can be written as
\begin{align}
  \langle \cdot \rangle_{\lambda\text{-odd}} =
  \frac{i N}{4} \frac{\sin^2\!\left( \frac{\pi\lambda}{2} \right)}{\pi\lambda}
  \left[
  \frac{1}{q''} - \frac{1}{q'} + \frac{1}{|q|}
  \left( \frac{|q''|}{q'} - \frac{|q'|}{q''} \right)
  \right] \ed
  \label{J0JpJmOdd}
\end{align}
In this case it is not important for us whether there is a contact term, because the $\lambda$ dependence of the odd structure (which is what we are interested in) cannot be changed by its presence. Indeed, this $\lambda$ dependence is determined by the last term in the square brackets, which cannot be a contact term.

\subsection{$\left< J^{(2)} J^{(2)} \right>$}

\begin{figure}
  \centering
  \includegraphics[width=0.8\textwidth]{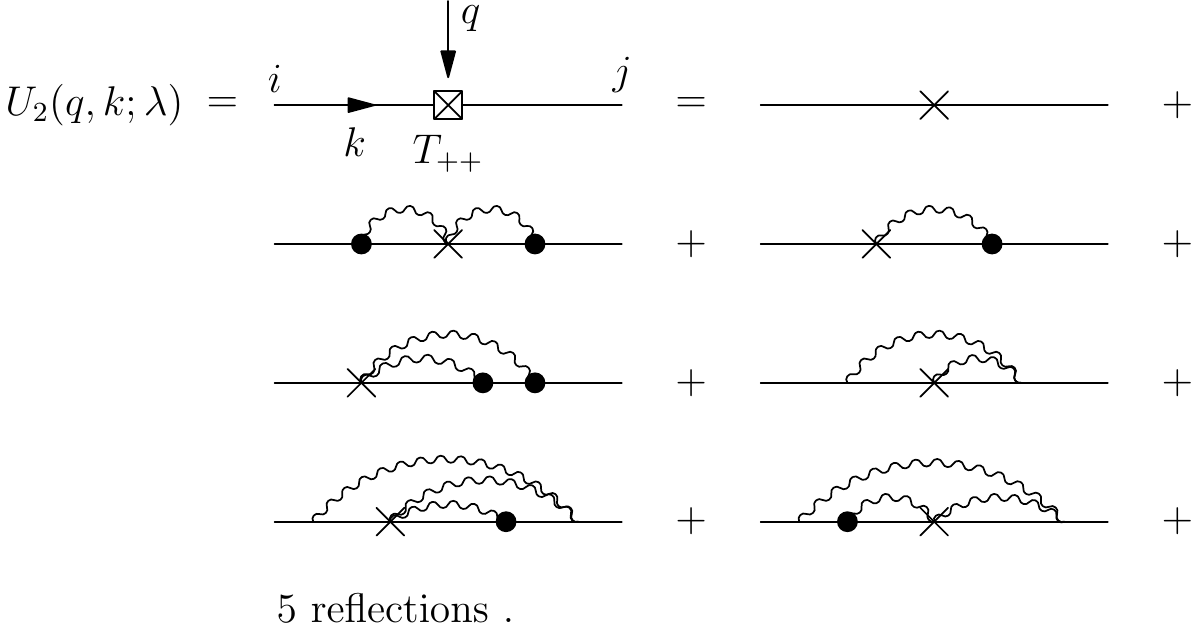}
  \caption{Diagrams of $\langle T_{++} \phi^\dagger \phi \rangle$ in which a gauge
  field is connected to $T_{++}$.}
  \label{fig:Tsqr}
\end{figure}

\begin{figure}
  \centering
  \includegraphics[width=0.8\textwidth]{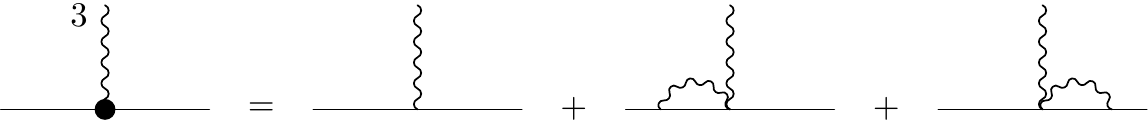}
  \caption{Shorthand notation for the exact vertex $\langle \phi^\dagger \phi A_+
  \rangle$.}
  \label{fig:Tshorthand}
\end{figure}

In this section we compute the correlator $\langle T_{--}(-q) T_{++} \rangle$ with $q^\pm = 0$. We again introduce the notation of a squared vertex $U_2$ for $T_{++}$, shown in figure \ref{fig:Tsqr} (similar to figure \ref{fig:Jp}). The filled-circle vertex is the exact vertex $\langle \phi^\dagger \phi A_+ \rangle$, shown in figure \ref{fig:Tshorthand}.

We find that
\begin{align}
  U_2(q,k;\lambda) &= \frac{k_s^2}{6 (k^+)^2} \left[
  3 \left(1+\lambda^2\right) k_s^2 +
  3 \lambda^2 k_3 (k_3 + q_3) +
  2 i \lambda \left(2+\lambda ^2\right) k_s q_3 \right]
  \delta_{i}^{j} \ed
  \label{U2}
\end{align}
The 2-point function of the stress-energy tensor is then given by figure \ref{fig:JmJp}, replacing $J_-$ with $T_{--}$ and $J_+$ with $T_{++}$. The exact $T_{--}$ vertex is given by $V_2$. Using \eqref{Vs} and \eqref{U2}, the result is
\begin{align}
  \langle  T_{--}(-q) T_{++} \rangle =
  \frac{N}{384 \pi \lambda} &\Big[
  3 i q_3^3 \left(1 - e^{2 i \hlambda \arctan (2 \Lambda')} \right)
  - 12 \lambda q_3^2 \Lambda
  \notag \\ &\quad
  - 24 i \lambda^2 q_3 \Lambda^2 +
  16 \lambda (1 + 2 \lambda^2) \Lambda^3
  \Big] \ed
  \label{TTall}
\end{align}
The result should contain a single parity-even conformal structure, up to
divergences and contact terms.  The divergent pieces in \eqref{TTall} can be
subtracted by counterterms of the form $\Lambda \partial_\mu g_{\nu\rho}
\partial^\mu g^{\nu\rho}$, $\Lambda^2 \epsilon_{\mu\nu\rho} g^{\rho\sigma}
\partial^\mu g^\nu_{\;\;\sigma}$, and $\Lambda^3 g_{\mu\nu} g^{\mu\nu}$, where
$g_{\mu \nu}$ is the linearized background metric that couples to the stress-tensor. The
finite $\lambda$-odd piece, proportional to $q_3^3$, can come from the contact term \cite{Closset:2012vp}
\begin{align}
  \langle T_{\mu\nu}(-q) T_{\rho\sigma} \rangle &\sim
  \left[ \epsilon_{\mu\rho\lambda} q^\lambda (q_\nu q_\sigma - q^2 \delta_{\nu\sigma})
  + (\mu\leftrightarrow\nu) \right] + (\rho\leftrightarrow\sigma) \ed
\end{align}

The remaining finite $\lambda$-even piece is
\begin{align}
  \frac{N |q_3|^3}{128} \frac{\sin(\pi \lambda )}{\pi\lambda} \ed
  \label{TTfinal}
\end{align}

\subsection{$\left< J^{(2)} J^{(1)} J^{(1)} \right>$}

In this section we compute the correlator  $\langle T_{--}(-q) J_+(-q') J_+(-q'') \rangle$, with all external momenta in the 3-direction. We expect two parity-even conformal structures, corresponding to the free bosonic and fermionic theories, and one parity-odd structure \cite{Giombi:2011rz}.

\begin{figure}
  \centering
  \includegraphics[width=1\textwidth]{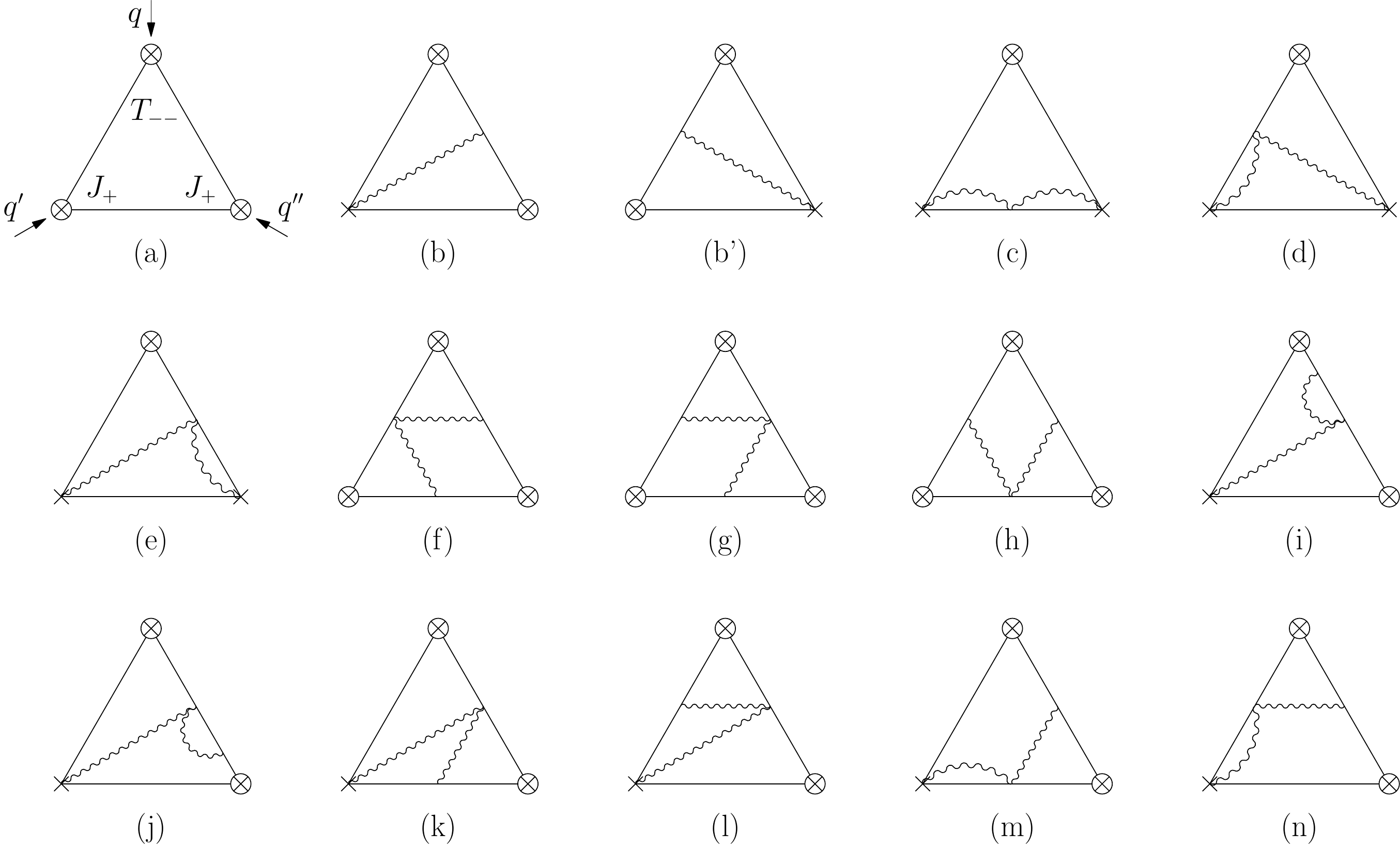}
  \caption{Diagrams contributing to $\langle T_{--}(-q) J_+(-q') J_+(-q'')
  \rangle$. In addition to these, there are reflections of (i)-(n) about the
  vertical axis.}
  \label{fig:TJJ}
\end{figure}

The diagrams are shown in figure \ref{fig:TJJ}.
As in the case of $\langle J^{(0)} J^{(1)} J^{(1)} \rangle$, all diagrams that contain the seagull vertex cancel in triples. We are left with $\mathrm{(a)}+\mathrm{(b)}+\mathrm{(b')}$. After integrating over the $p_3$ and angular directions, they are given by
\begin{align}
  \mathrm{(a)}  &= \frac{N}{16\pi}
  e^{2 i \hlambda \arctan{2\Lambda'}} \int_0^{\Lambda} dp_s
  \frac{\left(24 p_s^2 + q^2 + q'^2 + q''^2\right)}
  {(4p_s^2+q^2)(4p_s^2+q'^2)(4p_s^2+q''^2)} \times \notag \\
  &\quad
  \left[ 4p_s^2 + q'^2 \left( 1 - e^{-2 i \hlambda' \arctan(2y')}\right) \right]
  \left[ 4p_s^2 + q''^2 \left( 1- e^{-2 i \hlambda'' \arctan(2y'')}\right) \right]
  e^{-2 i \hlambda \arctan(2y)} \ec \notag \\
  \mathrm{(b)} &= \frac{i N}{8 \pi} e^{2 i \hlambda \arctan{2\Lambda'}}
  (q''-q) \int_0^{\Lambda} \frac{dp_s}{4p_s^2 + q^2}
  \left[ 4\lambda p_s + i q'' \left( 1 -  e^{-2 i \hlambda'' \arctan(2y'')} \right)
  \right] e^{-2 i \hlambda \arctan(2y)}
  \ec \notag \\
  \mathrm{(b')} &= \frac{i N}{8 \pi} e^{2 i \hlambda \arctan{2\Lambda'}}
  (q'-q) \int_0^{\Lambda} \frac{dp_s}{4p_s^2 + q^2}
  \left[ 4\lambda p_s + i q' \left( 1 -  e^{-2 i \hlambda' \arctan(2y')} \right)
  \right] e^{-2 i \hlambda \arctan(2y)} \ed
\end{align}

The integral obtained by summing over (a), (b) and (b') can be computed analytically, and it has a linear divergence: $\frac{3\Lambda}{8\pi}$. Subtracting the divergence\footnote{This divergence is subtracted by the counter-term $g^{--} \cA^+ \cA^+$, which is related by $SO(3)$ invariance to the term $g^{+-} \cA^+ \cA^-$ that we already used to subtract the divergence in $\langle J_+ J_- \rangle$ (see equation \eqref{JmJpraw}). However, our regularization preserves only $SO(2)$ invariance in the $x^1-x^2$ plane (as well as parity duality) under which these terms are not related. The two subtractions are therefore independent.} we obtain the 3-point function,
\begin{align}
  \langle T_{--}(-q) & J_+(-q') J_+(-q'') \rangle = -\frac{i N}{32 \pi \lambda}
  \frac{1}{q q' q''} \left[ e^{-\pi i \hlambda'}(q'^4-q'^2q''^2)
  + e^{-\pi i \hlambda''}(q''^4-q'^2q''^2) + \right.\nonumber\\
  &\quad \left. q'^2q''^2e^{-\pi i (\hlambda'+\hlambda'')}
  - e^{\pi i \hlambda}q^4 +  q' q'' (4q'^2 + 7 q' q'' + 4q''^2)
  - 12 \lambda^2 q' q'' q^2 \right] \ed
  \label{TJJall}
\end{align}
The $\lambda$-even part can be written as
\begin{align}
\left<\cdot\right>_{\lambda-\text{even}} &=
-\frac{N}{32\pi} \frac{\sin(\pi\lambda)}{\lambda}\left[ \left( \frac{|q|^3}{q' q''} + \frac{|q'|^3}{q q''} + \frac{|q''|^3}{q q'} \right) \cos^2\left( \frac{\pi\lambda}{2} \right) + \right. \notag \\
&\quad \left. \left( \frac{|q|^3}{q' q''} + \frac{|q'|^3}{q q''} + \frac{|q''|^3}{q q'} - 2\frac{q''|q'| + q'|q''|}{q} \right) \sin^2\left(\frac{\pi\lambda}{2}\right) \right] \ed \label{eq:TJJeven}
\end{align}
Here, the bosonic structure is the one multiplying $\cos^2\left( \frac{\pi\lambda}{2} \right)$, since it is the one that survives when taking $\lambda\to 0$. The fermionic structure multiplies $\sin^2\left(\frac{\pi\lambda}{2}\right)$, up to possible contact terms.

The $\lambda$-odd part of \eqref{TJJall} is
\begin{align}
  \left<\cdot\right>_{\lambda-\text{odd}} &=
  \frac{N i}{32\pi\lambda} \left[ 4( 3\lambda^2 + \cos(\pi\lambda) - 1) q + \sin^2(\pi\lambda) \frac{|q'||q''|+q' q''}{q} \right]\ed \label{eq:TJJodd}
\end{align}
The first term inside the brackets is a contact term, while the second term is the expected parity-odd structure.

\subsection{Correlators at the Critical Fixed Point}
\label{crit}

In this section we consider the planar correlation functions in the ``critical fixed point'' of the bosonic vector model with Chern-Simons interactions. This fixed point is reached by starting with the theory we discussed above, turning on the relevant ``double-trace'' deformation $\frac{\lambda_4}{2N}(\phi^{\dag}\phi)^2$, and flowing to the IR while tuning the IR scalar mass to zero \cite{Wilson:1971dc,Wilson:1973jj}. This is equivalent to adding an auxiliary field $\sigma$ with a $\sigma (\phi^{\dag} \phi)$ coupling, or performing a Legendre transform with respect to the operator $J^{(0)}$. Alternatively, one can start with the usual ``critical $U(N)$ model'' and couple it to the $U(N)_k$ Chern-Simons theory.

\begin{figure}
  \centering
  \includegraphics[width=0.4\textwidth]{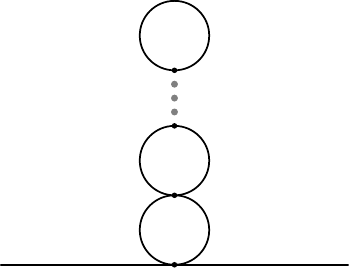}
  \caption{Corrections to the scalar propagator in the critical theory, in the
  planar limit. Gluon lines can run inside the scalar loops.}
  \label{fig:crit-tadpole}
\end{figure}

In the planar limit the effects of the ``double-trace'' deformation $(J^{(0)})^2$ are rather simple. First, the scalar propagator receives corrections involving a chain of scalar loops, connected by the $\lambda_4$ vertex, and ending in a tadpole (see figure \ref{fig:crit-tadpole}). In the planar limit one can add gluon lines inside each scalar loop in this chain, such that the scalar loop has the topology of a disk in double-line notation. These corrections are all power-law divergent and independent of the scalar momentum. They are subtracted with a mass counter-term $\phi^\dagger\phi$.

\begin{figure}
  \centering
  \includegraphics[width=0.8\textwidth]{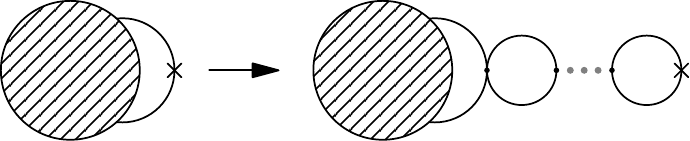}
  \caption{Corrections to an insertion of a ``single-trace'' operator in the
  critical theory, in the planar limit. Again, gluon lines can run inside the
  scalar loops.}
  \label{fig:crit-insertion}
\end{figure}

Second, in correlators of gauge-invariant operators each insertion can be connected to a similar chain of scalar loops that ends on the rest of the diagram, as shown in figure \ref{fig:crit-insertion}. In momentum space the scalar loops factorize, and we can use our previous results to sum over the chains of scalar loops.
For instance, using diagrammatics as above, or using conformal perturbation theory in the large $N$ limit, the 2-point functions in the presence of the double-trace deformation can be written as
\begin{align}
\langle J^{(s)} J^{(s')} \rangle_{\lambda_4} &= \langle J^{(s)} J^{(s')} \rangle_{\lambda_4=0} \notag\\
    &\quad -\frac{\lambda_4}{N} \langle J^{(s)} J^{(0)} \rangle_{\lambda_4=0} \left[\sum_{n=0}^{\infty}\left(-\frac{\lambda_4}{N}\langle J^{(0)} J^{(0)} \rangle_{\lambda_4=0}\right)^n \right] \langle J^{(0)} J^{(s')} \rangle_{\lambda_4=0} \ed
\end{align}
The second line vanishes unless $s=s'=0$, since $\langle J^{(s)} J^{(0)} \rangle_{\lambda_4=0}=0$ if $s\neq 0$ from conformal invariance. For $s=s'=0$ we can use the exact 2-point function of $J^{(0)}$ \eqref{eq:J0J0} to sum up the series,
\begin{align}
\langle J^{(0)}(-q) J^{(0)} \rangle_{\lambda_4} &= \langle J^{(0)}(-q) J^{(0)} \rangle_{\lambda_4=0}\sum_{n=0}^{\infty}\left( -\frac{\lambda_4}{N} \langle J^{(0)}(-q) J^{(0)} \rangle_{\lambda_4=0} \right)^n \notag\\
&= \frac{N}{\lambda_4} \frac{1}{1+\frac{4\pi\lambda}{\tan\left(\frac{\pi\lambda}{2}\right)} \frac{|q|}{\lambda_4} } \ed
\end{align}

To reach the IR fixed point we take $\lambda_4 \to \infty$, so we expand in $\frac{|q|}{\lambda_4} \ll 1$ and pick up the leading term (dropping the contact term $N/\lambda_4$). Defining the scalar operator of the critical fixed point to be $\tilde{J}^{(0)} \equiv \lambda_4 J^{(0)}$ we obtain that
\begin{align}
\langle \tilde{J}^{(0)}(-q) \tilde{J}^{(0)} \rangle^{\text{crit.}} = -N\frac{4\pi\lambda}{\tan\left(\frac{\pi\lambda}{2}\right)} |q| \ec
\label{J0J0crit}
\end{align}
which is the 2-point function of a primary operator of dimension 2 as expected.

Similarly, the 3-point functions with one scalar operator in the presence of the $\lambda_4$ deformation are given by
\begin{align}
\langle J^{(0)}(-q) J^{(s)} J^{(s')} \rangle_{\lambda_4} &= \langle J^{(0)}(-q) J^{(s)} J^{(s')} \rangle_{\lambda_4=0}\sum_{n=0}^{\infty}\left(-\frac{\lambda_4}{N} \langle J^{(0)} J^{(0)}(-q) \rangle \right)^n \notag \\
&\xrightarrow{\mathrm{IR}} \langle J^{(0)}(-q) J^{(s)} J^{(s')} \rangle_{\lambda_4=0} \frac{4\pi\lambda}{\tan\left(\frac{\pi\lambda}{2}\right)} \frac{|q|}{\lambda_4} \ed
\end{align}
In particular using \eqref{J0JpJmEven} (dropping the contact term) and \eqref{J0JpJmOdd} we obtain,
\begin{align}
  \langle \tilde{J}^{(0)}(-q) J_+(-q') J_-(-q'') \rangle_{\lambda\text{-even}}^{\text{crit.}} &= \frac{N}{2} \frac{\sin(\pi\lambda)}{\tan\left(\frac{\pi\lambda}{2}\right)} \left[
  \frac{|q|}{q'q''} + \frac{|q'|}{qq''} + \frac{|q''|}{qq'}
  \right]|q| \ec \label{cJ0JJeven} \\
  \langle \tilde{J}^{(0)}(-q) J_+(-q') J_-(-q'') \rangle_{\lambda\text{-odd}}^{\text{crit.}} &=
  \frac{i N}{2} \sin\!\left( \pi\lambda \right)
  \left[
  \frac{1}{q''} - \frac{1}{q'} + \frac{1}{|q|}
  \left( \frac{|q''|}{q'} - \frac{|q'|}{q''} \right)
  \right]|q| \ed
  \label{cJ0JJOdd}
\end{align}

By similar methods,
the correlator of three scalar operators at the critical fixed point evaluates to
\begin{align}
\langle \tilde{J}^{(0)}(-q) \tilde{J}^{(0)}(-q') \tilde{J}^{(0)}(-q'') \rangle^{\text{crit.}} \propto \langle J^{(0)}(-q) J^{(0)}(-q') J^{(0)}(-q'') \rangle_{\lambda_4=0} |q||q'||q''| \ed
\end{align}
The extra $|q| |q'| |q''|$ factor cancels all the momentum dependence of \eqref{eq:J0J0J0} so we get a constant, which is a pure contact term. We therefore obtain that up to contact terms
\begin{align}
\langle \tilde{J}^{(0)}(-q) \tilde{J}^{(0)}(-q') \tilde{J}^{(0)}(-q'') \rangle^{\text{crit.}} = 0 \ed
\label{eq:J0J0J0crit}
\end{align}
For $\lambda=0$ this is a standard result for the critical fixed point \cite{Petkou:1994ad}. The vanishing for all values of $\lambda$ is consistent with similar results for the theory of fermions coupled to Chern-Simons gauge fields, and for the analogous computations in Vasiliev's theory of gravity \cite{Giombi:2011kc}.

All correlators that do not involve a $J^{(0)}$ insertion remain unchanged at the critical fixed point
in the planar limit.

\section{Analysis of the Results}
\label{micro}

Let us summarize the $N$ and $\lambda$ dependence of the conformal structures that appear in the various correlators computed in section \ref{correlators}.

\paragraph{\underline{2-point functions:}}
\begin{align}
  \langle J^{(0)} J^{(0)} \rangle &= \frac{4 N \tan\left(\frac{\pi\lambda}{2}\right)}{\pi\lambda} \langle J^{(0)} J^{(0)} \rangle_{\text{bos.}} \ec \\
  \langle J^{(1)} J^{(1)} \rangle &= \frac{2 N \sin\left(\pi\lambda\right)}{\pi\lambda} \langle J^{(1)} J^{(1)} \rangle_{\text{bos.}} \ec \label{sumone} \\
  \langle J^{(2)} J^{(2)} \rangle &= \frac{2 N \sin\left(\pi\lambda\right)}{\pi\lambda} \langle J^{(2)} J^{(2)} \rangle_{\text{bos.}} \ed \label{sumtwo}
\end{align}
\paragraph{\underline{3-point functions:}}
\begin{align}
  \langle J^{(0)} J^{(0)} J^{(0)} \rangle &= \frac{4 N}{\pi\lambda}\left[ \frac{\tan\left(\frac{\pi\lambda}{2}\right)}{\cos^2\left(\frac{\pi\lambda}{2}\right)} - \frac{1}{4}\tan^3\left(\frac{\pi\lambda}{2}\right)\left(1+\frac{\lambda_6}{8\pi^2\lambda^2}\right) \right] \langle J^{(0)} J^{(0)} J^{(0)} \rangle_{\text{bos.}} \ec \\
  \langle J^{(1)} J^{(1)} J^{(0)} \rangle &= \frac{2 N \sin\left(\pi\lambda\right)}{\pi\lambda}
  \langle J^{(1)} J^{(1)} J^{(0)}\rangle_{\text{bos.}} + \frac{N \sin^2\left(\frac{\pi\lambda}{2}\right)}{\pi\lambda} \langle J^{(1)} J^{(1)} J^{(0)}\rangle_{\text{odd}} \ec \\
  \langle J^{(2)} J^{(1)} J^{(1)} \rangle &= \frac{2 N\sin(\pi\lambda)\cos^2\left(\frac{\pi\lambda}{2}\right)}{\pi\lambda} \langle J^{(2)} J^{(1)} J^{(1)}\rangle_{\text{bos.}} + \frac{2 N\sin(\pi\lambda)\sin^2\left(\frac{\pi\lambda}{2}\right)}{\pi\lambda}\langle J^{(2)} J^{(1)} J^{(1)}\rangle_{\text{fer.}} \notag \\
  &\quad + \frac{N \sin^2\left(\pi\lambda\right)}{\pi\lambda} \langle J^{(2)} J^{(1)} J^{(1)}\rangle_{\text{odd}} \ed
\end{align}
The correlators $\langle\cdot\rangle_{\text{bos.}}$, $\langle\cdot\rangle_{\text{fer.}}$ and $\langle\cdot\rangle_{\text{odd}}$ in the above equations were defined around equation \eqref{3pnt}\footnote{A correlator which involves $J^{(1)}$ in the free theory of a real field (boson or fermion) , is defined as $1/2$ of the same correlator for a complex field. The vector current and energy-momentum tensor of a free complex fermion are defined as $J_{\mu} = \bar{\psi}\gamma_{\mu}\psi$, $T_{\mu\nu} = \frac{1}{2}\bar{\psi}\gamma_{(\mu}\olra{\partial}_{\nu)}\psi$. With these normalizations the $2$-point functions of $J_\mu$ and $T_{\mu\nu}$ in the free boson and fermion theories agree.}. Note that we computed the correlators only for specific momenta, but using conformal invariance this is enough to determine the full correlators. One can verify that when $\lambda \to 0$ our correlators indeed go over to those of $2N$ free real bosons.

\subsection{Relation to the Results of Maldacena-Zhiboedov}

Let us map our microscopic couplings $N$, $\lambda$ to the parameters $\tilde{N}$, $\tilde{\lambda}$ of \cite{Maldacena:2012sf}, by comparing our correlators to the ones we reviewed in section \ref{mzreview}. In \cite{Maldacena:2012sf} the normalization of the energy-momentum tensor was chosen such that its 2-point function matches that of $\tilde{N}$ free real scalar fields. Our stress-tensor is canonically normalized at any value of $\lambda$, and since we can determine the overall normalization by requiring that in the $\lambda \to 0$ limit we have $2N$ free real scalar fields, we find from \eqref{sumtwo}
\begin{align}
  \tilde{N} = 2N \, \frac{\sin\left(\pi\lambda\right)}{\pi\lambda} \ed
  \label{Nt}
\end{align}
The result \eqref{sumone} is then also consistent with \eqref{2pntCurrents}, providing a consistency check on the assumption that the results of \cite{Maldacena:2012sf} hold also for the odd-spin currents. Our other results will provide additional consistency checks for this assumption. We interpret $\tilde{N}$ as the effective number of degrees of freedom in our theory, since $\langle T T \rangle$ is one way to define this number for a conformal theory; we find that it decreases as we increase the coupling, as expected, and that it goes to zero in the $\lambda \to 1$ limit (which is an infinite coupling limit using the Yang-Mills regularization to define the Chern-Simons coupling).

Next, we would like to compute $\tilde{\lambda}$, by comparing our correlators to the expressions of \cite{Maldacena:2012sf} for ``quasi-boson'' theories, written in \eqref{2pntCurrents}, \eqref{2pntQB} and \eqref{3pntQB}.
Let us define the normalized correlator
\begin{align}
  \left<J^{(s_1)} J^{(s_2)} J^{(s_3)}\right>^{\text{norm.}} \equiv
  \frac{\left<J^{(s_1)} J^{(s_2)} J^{(s_3)}\right>}
  {\sqrt{\left<J^{(s_1)} J^{(s_1)}\right>
  \left<J^{(s_2)} J^{(s_2)}\right>
  \left<J^{(s_3)} J^{(s_3)}\right>}}
  \ed
\end{align}
Comparing our results to equations \eqref{2pntCurrents}, \eqref{2pntQB} and \eqref{3pntQB} we have
\begin{align}
  \langle J^{(1)} J^{(1)} J^{(0)}\rangle^{\text{norm.}}_{\text{bos.}} \propto
  \left[ \tilde{N}(1+\tilde{\lambda}^2) \right]^{-1/2}
  \propto \left( \frac{ N
  \tan\left(\frac{\pi\lambda}{2}\right)}{\pi\lambda}
  \right)^{-1/2} \ed
\end{align}
Using \eqref{Nt}, we then find that
\begin{align}
  1 + \tilde{\lambda}^2 \propto
  \frac{1}{\cos^2 \left( \frac{\pi\lambda}{2} \right)} \ed
\end{align}
The proportionality constant can be fixed, for instance, by requiring that $\lambda=0$ corresponds to $\tilde{\lambda}=0$. The result is
\begin{align}
  \tilde{\lambda} &= \tan\left(\frac{\pi\lambda}{2}\right) \ec
  \label{lambdat}
\end{align}
where we have arbitrarily fixed the sign by assuming that $\lambda$ and $\tilde{\lambda}$ have the same sign.

One can now compute several other normalized correlators to check the results \eqref{Nt} and \eqref{lambdat}. We find
\begin{align}
  \langle J^{(1)} J^{(1)} J^{(0)}\rangle ^{\text{norm.}}_{\text{odd}}
  &\propto \left[
  \frac{\pi\lambda\tan\left(\frac{\pi\lambda}{2}\right)}{N}\right]^{1/2}
  \ec &
  \langle T J^{(1)} J^{(1)} \rangle ^{\text{norm.}}_{\text{bos.}}
  &\propto \left[
  \frac{\pi\lambda}{N \sin(\pi\lambda)}
  \right]^{1/2} \cos^2\left(\frac{\pi\lambda}{2}\right)
  \ec \notag \\
  \langle T J^{(1)} J^{(1)}\rangle ^{\text{norm.}}_{\text{fer.}}
  &\propto \left[
  \frac{\pi\lambda}{N \sin(\pi\lambda)}
  \right]^{1/2} \sin^2\left(\frac{\pi\lambda}{2}\right)
  \ec &
  \langle T J^{(1)} J^{(1)}\rangle ^{\text{norm.}}_{\text{odd}}
  &\propto \left[
  \frac{\pi\lambda \sin(\pi\lambda)}{N}
  \right]^{1/2}
  \ed
\end{align}
All of these results are precisely consistent with the results of \cite{Maldacena:2012sf}, using the values of ${\tilde N}$ and ${\tilde \lambda}$ computed above.

In the ``quasi-boson'' theory we also have the parameter $a_3$ which is the coefficient of the triple-trace term in $\partial\cdot J^{(4)}$. This term affects only $\langle  J^{(0)} J^{(0)} J^{(0)} \rangle $.
Expressing \eqref{eq:J0J0} and \eqref{eq:J0J0J0} in terms of $\tilde{\lambda}$ and $\tilde{N}$ we obtain
\begin{align}
\langle  J^{(0)} J^{(0)} \rangle  &\propto \tilde{N}(1+\tilde{\lambda}^2) \ec \\
\label{ourthree}
\langle  J^{(0)} J^{(0)} J^{(0)} \rangle  &\propto \tilde{N}(1+\tilde{\lambda}^2)\left[ 1 + \frac{\tilde{\lambda}^2}{4}\left( 3 - \frac{\lambda_6}{8\pi^2\lambda^2} \right) \right] \ed
\end{align}
From the 2-point function we see that to match with the normalization of \cite{Maldacena:2012sf} (see \eqref{2pntQB}) we must have $J^{(0)} = J^{(0)}_{\text{MZ}} (1+\tilde{\lambda}^2)$, and we then obtain from \eqref{ourthree}
\begin{align}
\langle  J^{(0)}_{\text{MZ}} J^{(0)}_{\text{MZ}} J^{(0)}_{\text{MZ}} \rangle  &\propto \frac{\tilde{N}}{(1+\tilde{\lambda}^2)^2}\left[ 1 + \frac{\tilde{\lambda}^2}{4}\left( 3 - \frac{\lambda_6}{8\pi^2\lambda^2} \right) \right] \ed
\end{align}
Therefore, matching to \eqref{3pntQB} we find
\begin{align}
a_3 \propto \frac{\tilde{\lambda}^2(1+\tilde{\lambda}^2)}{\tilde{N}^2}\left(3-\frac{\lambda_6}{8\pi^2\lambda^2}\right)
= \frac{\pi^2}{16 N^2 \cos^6(\frac{\pi\lambda}{2})} \left( 3 \lambda^2 - \frac{\lambda_6}{8\pi^2} \right)\ed
\end{align}
In particular, for $\lambda_6 = 24\pi^2\lambda^2$ we get $a_3=0$.

We can similarly map our results of section \ref{crit} for the critical fixed point to the results of \cite{Maldacena:2012sf} for ``quasi-fermion'' theories (see the discussion below \eqref{3pntQF}). Since $\langle TT \rangle$ is the same in the critical and non-critical theories, $\tilde{N}$ remains unchanged. To extract $\tlambda_{\text{qf}}$ of the quasi-fermionic theory consider the normalized correlator (see \eqref{J0J0crit}, \eqref{cJ0JJeven})
\begin{align}
\langle J^{(1)} J^{(1)} \tilde{J}^{(0)} \rangle_{\text{crit. bos.}}^{\text{norm.}} \propto \left(\frac{\tlambda_{\text{qf}}^2}{\tilde{N}(1+\tlambda_{\text{qf}}^2)}\right)^{1/2} \propto \left(\frac{\pi\lambda}{N\tan\left(\frac{\pi\lambda}{2}\right)}\right)^{1/2} \ed
\end{align}
Using $\eqref{Nt}$ in the above equation gives the relation
\begin{align}
\frac{\tlambda_{\text{qf}}^2}{1+\tlambda_{\text{qf}}^2} \propto \cos^2\left(\frac{\pi\lambda}{2}\right) \ed
\end{align}
Now, the $\lambda \to 0$ limit should correspond to $\tlambda_{\text{qf}} \to \infty$. This fixes the proportionality constant and we obtain
\begin{align}
\tlambda_{\text{qf}} = \cot\left(\frac{\pi\lambda}{2}\right) \ed
\end{align}
As we did above for the ``quasi-boson'' case, one can write down all the other normalized structures that we computed in section \ref{crit}, and verify that they are all consistent with the results of \cite{Maldacena:2012sf}.

So far we have matched the values of physical parameters to those of Maldacena and Zhiboedov using correlators at separated points. High-spin symmetry then determines all the 3-point functions at separated points.  As we saw in section \ref{sec:J1J1}, our theory also contains contact terms which contain physical information. It would be interesting to understand whether the high-spin symmetry constrains them, and to compare them to the fermionic theory.

\subsection{The Relation Between the Scalar and Fermionic Theories}

It was shown in \cite{Maldacena:2012sf} that the correlation functions of the ``quasi-boson'' theory, which is equal to the free bosonic theory as $\tilde\lambda \to 0$, become equal to those of (the Legendre transform of) the free fermionic theory as $\tilde\lambda \to \infty$. Similarly, the correlators of the ``quasi-fermion'' theory (which is equal to the free fermion theory as $\tilde\lambda_{\text{qf}}\to 0$) become those of the critical $O(N)$ (or $U(N)$) scalar model when $\tilde\lambda_{\text{qf}} \to \infty$.

Our results above imply that the limit of $\tilde\lambda \to \infty$ corresponds to $\lambda \to 1$, which is the maximal allowed coupling when regularizing the Chern-Simons theory with Yang-Mills terms.
In this limit our results should thus correspond to a Legendre transform of the theory of $N_\fer$ free complex fermions, for some value of $N_\fer$.
Let us determine this value by matching $\langle TT \rangle$ between the two theories (note that in \cite{Maldacena:2012sf}, the normalization is such that this correlator is the same for a free fermion and for a free boson).

In the free fermion theory (and in its Legendre transform), we have simply $\langle TT \rangle = 2N_\fer \langle TT \rangle_1$, where $\langle TT \rangle_1$ is the result for a single real free boson or fermion. In our bosonic theory we need to take the limit $\lambda \to 1$ while simultaneously taking $N \to \infty$. We can parameterize this limit by keeping $k_\YM = k-N$ fixed while taking $N \to \infty$. From \eqref{sumtwo} we then see that in this limit $\tilde{N} \to 2k_\YM$.
Thus, we expect our theory to go over to the Legendre transform of the theory of $N_\fer = k_{\YM}$ free complex fermions in this limit.
It is nice to see that this result is always an integer, although since we derived it only in the large $N$ limit with fixed $N/k$, it could be subject to shifts of order one (which must still give an integer).

Next, let us determine the Chern-Simons level of this fermionic theory when we move slightly away from the free fermion point. In our bosonic theory, using \eqref{lambdat} and taking the $\lambda\to 1$ limit as defined above, we see that $\tilde\lambda \simeq \frac{2}{\pi (1 - \lambda)}$.
In the fermionic $U(N_{\fer})$ theory at weak coupling, $\tilde\lambda_{\text{qf}}$
was normalized in \cite{Maldacena:2012sf} so that $\tilde\lambda_{\text{qf}} = \frac{\pi}{2} \frac{N_{\fer}}{k_{\fer}}$
(see equation (4.25) of \cite{Maldacena:2012sf}). Since 3-point functions of operators with $s>0$ are independent of the Legendre transform in the fermionic theory, we can use this result also for the Legendre-transformed theory, but first we must translate from this ``quasi-fermionic'' variable $\tilde\lambda_{\text{qf}}$ to the ``quasi-bosonic'' variable $\tilde{\lambda}$ used above. Matching the high-spin correlators \eqref{3pntQB}, \eqref{3pntQF} (which are independent of the Legendre transform in the large $N$ limit), we find that they are related by $\tilde\lambda = 1 / \tilde\lambda_{\text{qf}}$, and we then find that $k_{\fer} = k$ (up to a possible sign\footnote{This sign can be determined by comparing the signs
of the one-loop corrections to the three-point functions of currents in the bosonic and fermionic theories, but we will not do this here.}).
This implies that in the large $N$ limit, the theory of $N$ scalars coupled
to a $U(N)_k$ Chern-Simons theory is equivalent to the (Legendre transform of
the) theory of $(k-N)$ fermions coupled to a $U(k-N)_k$ Chern-Simons theory. If
we translate the Chern-Simons level to the one defined using the Yang-Mills
regularization, we obtain that the theory of $N$ scalars coupled to a
$U(N)_{k_{\YM}}$ Chern-Simons theory is equivalent in the large $N$ limit to
the (Legendre transform of the) theory of $k_{\YM}$ fermions coupled to a
$U(k_{\YM})_{N}$ Chern-Simons theory.

As a first consistency check on this statement, note that the Chern-Simons theories (without the matter) that we find on both sides of this relation are equivalent by level-rank duality \cite{Naculich:1990pa,Camperi:1990dk,Mlawer:1990uv} (this is true even at finite $N$, and certainly at large $N$). This is an important consistency check on the duality, since in the large $N$ limit, the computations of many objects in these theories (like the $S^3$-partition function or correlation functions of Wilson lines) are dominated by the Chern-Simons contributions, which scale as $N^2$, and these must agree for the duality to make sense. Moreover, the level-rank duality of Chern-Simons theories exchanges Wilson lines in symmetric representations with those in anti-symmetric representations, which meshes well with the exchange of scalars with fermions. It would be interesting to see if one could perhaps derive the scalar-fermion duality by integrating out the scalars in one theory and the fermions in the other theory (at least at large $N$), expressing the results as correlation functions of Wilson lines in the pure Chern-Simons theory (see \cite{Strassler:1992zr} and references therein), and seeing if these correlation functions are related by level-rank duality. It would be interesting to perform further tests of the duality, for instance by computing the effective potential on both sides, or by comparing the fractional contact term coefficients mentioned above.

The duality described above may be viewed as a large $N$ version of bosonization in three dimensions; a theory of fermions coupled to a Chern-Simons theory is described in a purely bosonic language (the theory of fermions
without any coupling to Chern-Simons is described here as the infinite coupling limit of a bosonic theory).
It is interesting to ask if this bosonization could be an exact equivalence also for finite values of $N$. Since the duality between the bosonic and fermionic theories exchanges weak and strong coupling, it is very difficult to test if this is true (given that we do not know how to perform exact computations at finite $N$)\footnote{In the bosonic theory at finite $N$, or in the Legendre transform of the fermionic theory, the classically marginal coupling $\lambda_6$ has a non-trivial beta function and should be taken to its fixed point. It was shown in \cite{Aharony:2011jz} that such an IR-stable fixed point exists at large $N$ and small $\lambda$, and it would be interesting to understand exactly when it exists.}. In two dimensional bosonization we know how to construct the fermion operators as solitons in the bosonic theory, and to prove the duality rigorously, but it is not clear how to do this in our case. In supersymmetric theories we can test similar dualities by comparing moduli spaces, chiral rings, and so on, but we do not have this privilege in our case. It is interesting to note that the transformation of the Chern-Simons group in our theory is the same as that of the Seiberg-like duality found in \cite{Giveon:2008zn} for ${\cal N}=2$ supersymmetric Chern-Simons-matter theories (up to a shift by the number of flavors $N_f$ which is not visible in our large $N$ limit); of course in that case both sides contain both scalars and fermions coupled to the Chern-Simons theory, while in our case we have only scalars on one side and only fermions on the other side. It is interesting to ask\footnote{We thank D. Kutasov for suggesting this.} if the scalar-fermion duality could perhaps be derived by flowing from the supersymmetric duality, in which case we could confirm its validity for finite $N$ (since the supersymmetric duality is believed to be valid also at finite $N$). 

One test that we can perform at finite $N$ involves deformations of our theories. In this discussion, for simplicity, we use the definition of the Chern-Simons couplings using the Yang-Mills regularization. First, note that the fermionic theory we discussed cannot really be at level $k_{\fer}=N$, since in the presence of one flavor the level must be half-integer for the theory to be gauge-invariant \cite{Niemi:1983rq,Redlich:1983kn,Redlich:1983dv}. Let us assume that the correct level is $k_{\fer}=N-\frac{1}{2}$. Now, let us start with the fermionic theory coupled to a $U(k_{\YM})_{N-1/2}$ Chern-Simons theory, and deform it by a mass term to the fermions, $M \bar\psi^a \psi^a$. At scales below $M$ we can integrate out the fermions, and remain with a pure Chern-Simons theory, whose level depends on the sign of $M$ \cite{Niemi:1983rq,Redlich:1983kn,Redlich:1983dv}. For one sign we end up with a (topological) $U(k_{\YM})_N$ Chern-Simons theory, and for the other sign with a $U(k_{\YM})_{N-1}$ Chern-Simons theory.

For the duality to be valid also at finite $N$, we must obtain the same low-energy theory also on the bosonic side. On that side we start with the critical bosonic theory, which can be viewed as the deformation of a theory of scalars coupled to $U(N)_{k_{\YM}}$ by $\sigma \phi^{\dagger} \phi$, and deform it by $M \sigma$. The auxiliary field $\sigma$ now serves as a Lagrange multiplier, enforcing $\phi^{\dagger} \phi = - M$, and we need to understand the behavior of the bosonic theory with this constraint. For large values of $N$ and any value of $\lambda$, one can show that for positive $M$ the bosonic theory has a stable vacuum with unbroken $U(N)$, in which the scalars are massive, so that at low energies we obtain the $U(N)_{k_{\YM}}$ pure Chern-Simons theory. For negative $M$ there is no such vacuum, but there is an alternative vacuum in which one of the scalars condenses, and the gauge symmetry is broken to $U(N-1)$. In this case we find at weak coupling that the other $(2N-1)$ gauge bosons become massive and dynamical (by swallowing scalar fields), as does the remaining real scalar, so we obtain at low energies the $U(N-1)_{k_{\YM}}$ pure Chern-Simons theory. These computations are done assuming that $N$ is large, and in some cases also that the coupling constant is small, but since the low-energy theory is a topological theory labeled by discrete parameters, we expect to find the same low-energy theory for any $N$ and $k$. The two low-energy theories that we find here, for the two signs of $M$, are precisely equivalent (by level-rank duality) to the two theories that we found in the fermionic case, thus providing weak evidence for the validity of the bosonization also at finite $N$.

It is natural to generalize to the case where we have $N_f$ flavors of massless scalars/fermions in the fundamental representation of some $U(N)$ group. The computations of the large $N$ correlation functions that we computed above for this case are a straightforward generalization of our computations in the previous sections, though now we have $N_f^2$ operators of each spin, so it is not a priori obvious if the results of \cite{Maldacena:2012sf} can be applied\footnote{There are also new correlation functions that can appear when $N_f > 1$, such as terms proportional to $f^{ABC}$ in $\langle J^{(1)A} J^{(1)B} J^{(1)C} \rangle$ (see \cite{Giombi:2011rz}), and it would be interesting to compute them and to check if they are consistent with the duality.}. In any case, the natural conjecture following from the discussion of the previous paragraph is that (using the Yang-Mills definition for the Chern-Simons coupling) the theory of $N_f\cdot N$ scalars coupled to the $U(N)_{k_{\YM}}$ Chern-Simons theory is equivalent (up to a Legendre transform) to the theory of $N_f\cdot k_{\YM}$ fermions coupled to the $U(k_{\YM})_{N-N_f/2}$ Chern-Simons theory. Translating back to our definition of the coupling, we map the theory of $N_f$ scalars in the fundamental representation coupled to $U(N)_k$ to the theory of $N_f$ fermions in the fundamental representation coupled to $U(k-N)_{k-N_f/2}$. This is very similar to the supersymmetric dualities of \cite{Giveon:2008zn}. One difference is an overall shift in $k$ by $N_f/2$, which is probably related to the one-loop contributions of the fermionic fields to $k$. Another difference is taking $N_f/2 \to N_f$, which is related to the fact that in \cite{Giveon:2008zn} $N_f$ was the number of fundamental chiral multiplets, and also the number of anti-fundamental chiral multiplets, so the overall number of flavor fermions was doubled. A final difference is that in the supersymmetric case the sign of $k$ changes under the duality; above we did not fix this sign, and it is plausible that also in the scalar-fermion
duality one of the sides should have a negative Chern-Simons coupling (or equivalently, that if we keep positive Chern-Simons couplings, then the two sides are related by a parity transformation).

Note that the results we present here are much stronger than most previous results on bosonization in three dimensions (see, for instance, \cite{Deser:1988zm,Burgess:1994tm,Fradkin:1994tt,Banerjee:1995ry,Barci:1995iy,Banerjee:1996qu,Ghosh:1998bv,Barci:1998zd}), which claimed that the low-energy limit of the theory of massive fermions coupling to a gauge field is given by a Chern-Simons theory (and had non-local bosonic actions at higher energies). In our case we claim that for massless scalars/fermions we have an exact equivalence of conformal field theories. For the special case of $N=1$ and $k=1$, our duality seems very similar to the duality studied in
\cite{Shaji:1990is,Paul:1990vw,Shankar:1991mg}; it would be interesting to understand this better, and to see if the methods of these papers can be used to study our duality more generally.

\subsection{Comments on the Thermal Free Energy}

As described above, the duality is completely consistent with all correlation function computations done to date; preliminary computations of exact planar 2-point and 3-point correlators in the fermionic theory are also consistent with the duality presented in the previous subsection \cite{GY}.
However, the duality is not consistent (already at large $N$) with the form of the thermal free energy of the fermionic theory, computed in \cite{Giombi:2011kc}, as this form suggests a different relation between the bosonic and fermionic theories (as noted in \cite{Giombi:2011kc,Maldacena:2012sf}). It is also not consistent with the thermal free energy of the scalar theory \cite{us,Shiraz} if one computes it by similar methods, as this gives a result that does not even go to zero when $\lambda \to 1$. This is problematic independently of the duality, since we saw above that
$\langle T_{\mu \nu} T_{\rho \sigma} \rangle$ vanishes in this limit, so the number of degrees of freedom should go to zero, and this is not visible in the naive computation of the thermal free energy.

Our computations of the correlation functions pass many consistency checks; in particular they agree with the results of \cite{Maldacena:2012sf} which were computed by completely different methods. On the other hand, there are so far no consistency checks for the thermal free energy computations.
Thus, we claim that the existing computations of the thermal free energy are not correct. One possible
problem is the light-cone gauge which these computations use. This is defined by an analytic continuation from Minkowski space, and it is plausible that such a continuation gives correct results for correlation functions (which are analytic in the momenta), but not for the Euclidean partition function compactified
on a circle.

Another possible problem involves the large $N$ limit.
The computations in question are performed in the limit of large $N$ and large volume, and they take the fermions to be anti-periodic on the Euclidean thermal circle, and the scalars to be periodic on the circle, namely they assume that the holonomy of the Chern-Simons gauge field around the Euclidean thermal circle is trivial ($A_0=0$). This assumption is
expected to be correct at very high temperatures. It is also valid if we take the large volume limit first, so that the dynamics of the zero mode of the holonomy is decoupled. However, it is not clear if it is valid when we take the large $N$ limit first. In particular,
as discussed in \cite{Shenker:2011zf}, in vector models coupled to Chern-Simons theories, the holonomy becomes trivial (at least on $S^2$) only for very large temperatures $T$, obeying $V T^2 \gg N$ (where $V$ is the spatial volume). In the free theory this was explicitly checked in \cite{Shenker:2011zf}, where it was found that for temperatures that do not scale with $N$ the holonomy is actually uniformly spread out on the thermal circle in the large $N$ limit, leading to a vanishing free energy at order $N$. We expect this to remain true also for finite $\lambda$.

The computations of the thermal free energy in the Chern-Simons-matter theories in \cite{Giombi:2011kc,us,Shiraz} use the standard 't Hooft limit, where only planar (disk) diagrams are kept. This is valid when we take the large $N$ limit first, keeping everything else (like the volume or temperature) fixed, and take any other limits (like large volume) later. As described above, if we do this for the theory on $S^2$ we land in the low-temperature phase, where the free energy vanishes at order $N$, so it does not give any useful comparisons between the bosonic and fermionic theories. (On other manifolds, like higher genus Riemann surfaces, the situation is more subtle since the Chern-Simons theory has many degenerate ground states \cite{newShenker}, and we will not discuss this case here.) In order to get a non-vanishing free energy at order $N$ we need to be in the high temperature phase, with $V T^2 \gg N$, but then it is not obvious that the standard 't Hooft large $N$ expansion applies.
This expansion is particularly subtle in our case, in which the leading order term, of order $N^2$, is given by a topological theory.
For instance, one may worry that for any $\lambda > 0$ connected diagrams with $n$ scalar loops could scale as $n$ powers of the volume, since in the pure Chern-Simons theory the correlators of $n$ Wilson lines are independent of their positions, and this power of $(V T^2)^n$ could overcome the suppression by $N^2/N^n$ in the high temperature phase\footnote{We thank S. Minwalla for suggesting that IR divergences may be responsible for the failure of the naive computation.}.
It would be interesting to try to fix the thermal free energy computation, and to use it to test the duality. Similarly, it would be interesting if (as in supersymmetric theories \cite{Kapustin:2009kz}) one could find a way to compute exactly the free energy of our theories on a Euclidean $S^3$ (at least in the large $N$ limit), as this could provide another useful test of the duality; this free energy for our theories at weak coupling may be found in \cite{Klebanov:2011gs}.

\subsection*{Acknowledgments}
\label{s:acks}

It is a pleasure to thank Simone Giombi, Amit Giveon, Nissan Itzhaki, Zohar Komargodski, David Kutasov, Juan Maldacena, Eliezer Rabinovici, Sandip Trivedi, Spenta Wadia, Xi Yin, and especially Shiraz Minwalla for many useful discussions. We especially thank Simone Giombi, Juan Maldacena, and Shiraz Minwalla for comments on a preliminary draft of this paper.
This work was supported in part by an Israel Science Foundation center for excellence grant, by the German-Israeli Foundation (GIF) for Scientific Research and Development, and by the Minerva foundation with funding from the Federal German Ministry for Education and Research.


\end{document}